\documentclass[12pt,prd,tightenlines,preprint]{revtex4-2}

\usepackage{amsmath,amssymb,color,makeidx,url,wrapfig}
\usepackage[pdftex]{graphicx}
\graphicspath{{./figures/}}

\DeclareMathOperator{\Var}{Var}
\DeclareMathOperator{\Cov}{Cov}

\begin{document}

\preprint{FERMILAB-PUB-23-286-T}

\title{Light Scalar Meson and Decay Constant in SU(3) Gauge Theory with Eight Dynamical Flavors}

\author{R.~C.~Brower}
\affiliation{Department of Physics and Center for Computational Science, Boston University, Boston, Massachusetts 02215, USA}
\author{C.~Culver}
\affiliation{Department of Mathematical Sciences, University of Liverpool, Liverpool L69 7ZL, UK}
\author{K.~K.~Cushman}
\affiliation{Department of Physics, Sloane Laboratory, Yale University, New Haven, Connecticut 06520, USA}
\author{G.~T.~Fleming}\email{gfleming@fnal.gov}
\affiliation{Department of Physics, Sloane Laboratory, Yale University, New Haven, Connecticut 06520, USA}
\affiliation{Theoretical Physics Division, Fermilab, Batavia, IL 60510, USA}
\author{A.~Gasbarro}
\affiliation{AEC Institute for Theoretical Physics, University of Bern, 3012 Bern, CH}
\author{A.~Hasenfratz}
\affiliation{Department of Physics, University of Colorado, Boulder, Colorado 80309, USA}
\author{J.~Ingoldby}
\affiliation{Abdus Salam International Centre for Theoretical Physics, Strada Costiera 11, 34151, Trieste, Italy}
\author{X.~Y.~Jin}
\affiliation{Computational Science Division, Argonne National Laboratory, Argonne, Illinois 60439, USA}
\author{E.~T.~Neil}
\affiliation{Department of Physics, University of Colorado, Boulder, Colorado 80309, USA}
\author{J.~C.~Osborn}
\affiliation{Computational Science Division, Argonne National Laboratory, Argonne, Illinois 60439, USA}
\author{E.~Owen}
\affiliation{Department of Physics and Center for Computational Science, Boston University, Boston, Massachusetts
02215, USA}
\author{C.~Rebbi}
\affiliation{Department of Physics and Center for Computational Science, Boston University, Boston, Massachusetts 02215, USA}
\author{E.~Rinaldi}
\affiliation{Interdisciplinary Theoretical and Mathematical Sciences Program (iTHEMS), RIKEN, 2-1 Hirosawa, Wako, Saitama 351-0198, Japan}
\author{D.~Schaich}
\affiliation{Department of Mathematical Sciences, University of Liverpool, Liverpool L69 7ZL, UK}
\author{P.~Vranas}
\affiliation{Physical and Life Sciences Division, Lawrence Livermore National Laboratory, Livermore, California 94550, USA}
\affiliation{Nuclear Science Division, Lawrence Berkeley National Laboratory, Berkeley, California 94720, USA}
\author{E.~Weinberg}
\affiliation{Department of Physics and Center for Computational Science, Boston University, Boston, Massachusetts 02215, USA}
\affiliation{NVIDIA Corporation, Santa Clara, California 95050, USA}
\author{O.~Witzel}
\affiliation{Center for Particle Physics Siegen (CPPS), Theoretische Physik 1, Naturwissenschaftlich-Technische Fakult\"at, Universit\"at Siegen, 57068 Siegen, Germany}

\collaboration{Lattice Strong Dynamics (LSD) Collaboration}
\noaffiliation


\begin{abstract}
The SU(3) gauge theory with $N_f=8$ nearly massless Dirac fermions has long been
of theoretical and phenomenological interest due to the near-conformality arising
from its proximity to
the conformal window. 
One particularly interesting feature is the emergence of a relatively light, stable
flavor-singlet scalar meson $\sigma$ $(J^{PC}=0^{++})$ in contrast to the $N_f=2$ theory QCD.
In this work, we study the finite-volume dependence of the $\sigma$ meson correlation function computed in lattice gauge theory
and determine the $\sigma$ meson mass and decay constant extrapolated
to the infinite-volume limit. We also determine
the infinite volume mass and decay constant of the flavor-nonsinglet scalar meson $a_0$.
\end{abstract}


\maketitle

\section{\label{sec:intro}Introduction}

SU(3) gauge theory with $N_f$ flavors of massless Dirac fermions has
a conformal window for $N_{fc} \le N_f \le 16$
\cite{Caswell:1974gg, Banks:1981nn}.  See \cite{Fleming:2008gy} for a review
of the early history of constraining $N_{fc}$
and see \cite{Drach:2020qpj} for the most recent review.
While it is not known with certainty whether
the massless $N_f=8$ theory is inside or outside the conformal window
\cite{Ishikawa:2015iwa, Fodor:2018uih, Fodor:2019ypi},
our collaboration has previously published results
\cite{LatticeStrongDynamics:2012nz, LatticeStrongDynamics:2020uwo}
indicating the massless $N_f=10$ theory is likely
inside the conformal window and for the rest of this paper we will assume the
massless $N_f=8$ theory is very close to the edge of the conformal window.
If it is inside the conformal window
it is most likely a very strongly coupled CFT \cite{Hasenfratz:2022qan}.
If it is outside the conformal window, spontaneous chiral symmetry breaking
and confinement produce
massless Nambu-Goldstone bosons and a spectrum of other hadronic states which may be
different relative to QCD due to the proximity of the conformal window.

The continuum SU(3) gauge theory with $N_f=8$ Dirac fermions
with small vector-like mass terms is not an IR conformal theory.
The small mass terms explicitly break chiral symmetry,
confinement
occurs and a massive spectrum of hadronic states is generated.
Another scenario may be possible at stronger lattice coupling
but we don't consider that here \cite{Hasenfratz:2022qan}.

In our previous papers \cite{LatticeStrongDynamics:2014nmn,
LatticeStrongDynamics:2018hun, LatticeStrongDynamics:2021gmp},
we identified two specific features of the low-energy spectrum which were different
from QCD.  First, the pion decay constant $F_\pi$ strongly depends on the fermion mass
unlike QCD where $F_\pi$ is nearly constant with a small, linear correction
in the fermion mass. Second, the flavor-singlet scalar meson $\sigma$
$(J^{PC}=0^{++})$ has a light mass $M_\sigma < 1.5\ M_\pi$
in the fermion mass region where we compute it,
well below the energy threshold for decay to two pions,
whereas in QCD it is somewhat heavier $M_\sigma > 1.9\ M_\pi$
\cite{Kunihiro:2003yj, Wakayama:2014gpa, Howarth:2015caa, Briceno:2016mjc,
Briceno:2017max, Briceno:2017qmb, Guo:2018zss, Rodas:2023gma, Rodas:2023twk},
remaining just below decay threshold across
in an equivalent fermion mass range.  The QCD picture is somewhat consistent
with our earlier $N_f=4$ calculation \cite{LatticeStrongDynamics:2018hun}.
However,  we also identified several features of the $N_f=8$ theory which appeared
similar to QCD calculations in an equivalent range of fermion masses: the ratios
$M_\rho / F_\pi$ and $M_\text{nucleon} / F_\pi$, the decay constants $F_\pi$, $F_\rho$
and $F_{a_1}$ appear consistent with QCD KSRF relations
\cite{Kawarabayashi:1966kd, Riazuddin:1966sw}
and the $I=2$ $\pi\pi$
scattering length $a_{\pi\pi}$ appears to agree with QCD.

In this paper, we focus larger volume calculations at the various fermion masses
(App.~\ref{app:ensembles})
which will allow us to extrapolate our results to the infinite volume limit,
removing one potential source of systematic error (Sec.~\ref{sub:finite_volume}).
We also introduce
an improved method for analyzing $\sigma$ meson correlation functions
with a new subtraction scheme in the rest frame combined with simultaneous
analysis in several moving frames
(Secs.~\ref{sub:model_function} and \ref{sub:dispersion_relation}).
We rely heavily on the method of Bayesian
model averaging \cite{Jay:2020jkz} (Sec.~\ref{sec:model_avg}).
We present new calculations of the flavor-singlet
scalar decay constant $F_S$ and the flavor-nonsinglet scalar meson $a_0$ mass
and decay constant $F_{a_0}$ (Sec.~\ref{sub:decay_constants}). We also
comment briefly on the chiral condensate and its contribution
to the Gell-Mann-Oakes-Renner
(GMOR) relation and its generalizations (Sec.~\ref{sub:GMOR_ratio}).

As described in our earlier paper \cite{LatticeStrongDynamics:2018hun}
we have chosen the bare lattice parameter $\beta=4.8$ such that the lattice spacing
$a$ is as coarse as possible given our current action, so that we can get as close
to the chiral limit $a m_q \to 0$ as possible with available computing resources.
We are working on calculations
at $a m_q=0.00056$ which may provide further insights in the near future.
If the massless $N_f=8$ theory is conformal and sufficiently strongly coupled
\cite{Hasenfratz:2022qan} then it is likely a new lattice action that allows
for even coarser lattice spacings will be necessary to make further progress.

Phenomenologically, theories that exhibit approximate conformal behavior
at strong coupling  are anticipated to produce large anomalous dimensions
over a wide interval of scales, which can make them attractive as candidate
composite Higgs models \cite{Panico:2015jxa, Ferretti:2016upr, Witzel:2019jbe,
Cacciapaglia:2019bqz, Cacciapaglia:2020kgq}.
In particular, the $\text{SU}(3)\,\,N_f=8$ theory has been used to build
composite Higgs models
in \cite{Vecchi:2015fma, Appelquist:2020bqj, Appelquist:2022qgl}.
The construction of a low energy EFT for the lightest composites,
to which the rest of the Standard Model can be coupled is a crucial intermediate step
in the creation of these models.
In a separate letter \cite{LatticeStrongDynamics:2023uzj}, we discuss various
effective models
that can be fit to our data.
\section{\label{sec:2pt}Staggered two-point correlation function construction and modeling}

\subsection{\label{sub:2pt_const}Staggered two-point correlation function construction}

The continuum $N_f=8$ theory is approximated on a finite lattice by an SU(2)-doublet of staggered fermion
fields $(\chi_1 \ \chi_2)$ that carry only an SU(3) color index at each lattice site.
Each component of the doublet represents
four non-degenerate Dirac fermion ``tastes'' with spin and taste degrees of freedom spread out over $2^4$
sites of local hypercubes.  In the continuum limit where the bare gauge coupling $g_0^2 \to 0$, these tastes
become degenerate and equivalent to four
Dirac flavors.  Hence the doublet of staggered fields becomes a degenerate $N_f=8$ theory in this continuum limit.
Staggered fermions have a remnant of chiral symmetry that can lead to $(N_f/4)^2$ Nambu--Goldstone bosons
when taking the massless chiral limit at finite lattice spacing, assuming the chiral symmetry
is spontaneously broken by the gauge interactions.  However, to recover the full flavor symmetry
it is essential to take the $g_0^2 \to 0$ continuum limit prior to the $m_q \to 0$ chiral limit.

In general, a staggered meson two-point correlaton function where source and sink operators have the same
quantum numbers $Q$ is (schematically)
\begin{equation}
C_Q(\vec{p},|t-t_0|) = \left\langle
\sum_{\vec{x}} e^{i \vec{p} \cdot (\vec{x}-\vec{x}_0)}
\overline{\chi}(\vec{x}+\vec{\delta}^\prime,t) \Gamma_Q(\vec{x},\vec{\delta}^\prime) \tau \chi(\vec{x},t)
\ \overline{\chi}(\vec{x}_0+\vec{\delta},t_0) \Gamma_Q(\vec{x}_0,\vec{\delta}) \tau \chi(\vec{x},t)
\right\rangle
\end{equation}
where $\Gamma_Q(\vec{x},\vec{\delta})$ are phases which refer to the spin-taste structure of the interpolating
operators with quantum numbers $Q$, and $\tau$ is either an SU(2) generator for a non-singlet correlator
or a $2 \times 2$ identity matrix for a singlet correlator under the SU(2) staggered doublet symmetry.  There are various phase conventions possible,
one common choice is \cite{Daniel:1987aa}.
Not shown are gauge matrices required to make the whole thing gauge-invariant, \textit{e.g.}\ connecting
sites $(\vec{x},t)$ and $(\vec{x}+\vec{\delta}^\prime,t)$. Also, translation invariance of the ensemble
average $\langle \ \cdot \ \rangle$ guarantees the correlation function depends only on the distance $|t-t_0|$
and not the source position $(\vec{x}_0,t_0)$.

In our earlier paper \cite{LatticeStrongDynamics:2018hun}, the LSD Collaboration constructed correlation functions with $\vec{p}=0$
for local and point-split operators.  In this study, we focused on constructing singlet and
non-singlet correlation functions of local operators at several different momenta $\vec{p}$
with much higher statistics.  On each gauge configuration, we generate a unique set of $N$ random source points
$(\vec{x}_0, t_0)_n$ and construct a primitive staggered meson ``connected'' correlator
\begin{equation}
C(\vec{x},t) = \frac{1}{N} \sum_n \mathcal{T}_n \ \text{Tr}_\text{color} \left[
G_F(\vec{x}_0,t_0;\vec{x},t) \ G_F^\dagger(\vec{x},t;\vec{x}_0,t_0)
\right]
\end{equation}
where $G_F(\vec{x}_0,t_0;\vec{x},t)$ is a $3 \times 3$ color matrix of single staggered fermion propagator
from the site $(\vec{x}_0,t_0)$ to the site $(\vec{x},t)$ and $\mathcal{T}_n$ represents the translation
of the $n$-th source location $(\vec{x}_0,t_0)_n$ to the origin $(\vec{0},0)$. Then, we record the value
of this averaged primitive correlator for every sink point $(\vec{x},t)$ in the lattice volume. We refer
to this as a connected correlator because valence fermion lines connect the source and sink points.  With
post-processing we can project this primitive correlator into eight different non-singlet staggered meson
quantum number channels of different momenta $\vec{p}$ using Fourier transform
\begin{equation}
\label{eq:nonsinglet_meson}
C_Q(\vec{p},t) = \sum_{\vec{x}} e^{i \vec{p} \cdot \vec{x}} \ C(\vec{x},t) \ \phi_Q(\vec{x}) .
\end{equation}
For example, if we choose the phase $\phi_Q(\vec{x}) = 1$, we get the correlation function for the $\pi_5$ meson,
which is the pseudo-Nambu--Goldstone boson.

Also in our earlier work, we explained in detail how we construct a ``disconnected'' correlator, where valence
fermion lines do not connect the source and sink points
\begin{equation}
D(\vec{p}, |t-t_0|) = \sum_{\vec{x}} \sum_{\vec{x}_0} e^{i(\vec{x}-\vec{x}_0) \cdot \vec{p}}
\ \text{Tr} \left[ G_F(\vec{x}_0,t_0;\vec{x}_0,t_0) \right]
\ \text{Tr} \left[ G_F(\vec{x},t;\vec{x},t) \right]
\end{equation}
using a diluted noisy estimator to compute the trace at each site on the lattice for each gauge configuration,
which is again recorded as a single value per site in the lattice volume.  With post-processing, we can compute
the disconnected correlator for any spatial momentum $\vec{p}$ using FFT and the fast convolution algorithm
\begin{equation}
\widetilde{O}(\vec{p},\omega) = \sum_{\vec{x}, t} e^{i (\vec{p} \cdot \vec{x} + \omega t)}
\ \text{Tr} \left[ G_F(\vec{x},t;\vec{x},t) \right]
\end{equation}
\begin{equation}
D(\vec{p},t) = \sum_\omega e^{-i \omega t} \left| \widetilde{O}(\vec{p},\omega) \right|^2
\end{equation}
where the result is automatically invariant under any lattice translation. In an $N_f$-flavor theory,
the flavor-singlet scalar correlator for the $\sigma$ meson is then
\begin{equation}
\label{eq:isosinglet}
C_\sigma(\vec{p}, t) = \left( \frac{N_f}{4} \right)^2 D(\vec{p}, t) - \frac{N_f}{4} C_{a_{0,1}}(\vec{p},t)
\end{equation}
and where $C_{a_{0,1}}(\vec{p},t)$ is the flavor-nonsinglet scalar meson correlator
constructed from Eq.~(\ref{eq:nonsinglet_meson}) with the appropriate choice of phases.
Note this normalization is different from \cite{LatticeStrongDynamics:2018hun} where we dropped
an overall factor of $N_f/4$.

Regarding the naming convention of mesons, we note that the continuum SU(8) flavor representation
is broken by lattice artifacts to a subgroup SU(2) $\times$ taste, a discrete subgroup of SU(4).
Meson names will follow the PDG convention for two-flavor mesons: $\pi$,
$a_0$, $\rho$, \dots plus an additional subscript to indicate the representation under the discrete
taste group: $\pi_5$, $a_{0,1}$, $\rho_i$, \dots There is only one scalar meson which is a singlet
over the whole flavor group which we name $\sigma$ and no subscript is required.
The effects of taste-breaking were discussed previously \cite{LatticeStrongDynamics:2018hun}
and we will not expand on it, so the taste index
not play a significant role here with one important exception.  In the continuum two-flavor theory,
the decay $a_0 \to \pi \pi$ is forbidden by isospin symmetry.  However, in our staggered $N_f=8$
theory, the decay $a_{0,1} \to \pi_5 \pi_5$ is allowed because the $\pi_5$ and the $a_{0,1}$
are not in the same SU(2) flavor subgroup, as indicated by the different taste indices.
It is analogous to the decay of $a_0 \to K K$ in continuum three-flavor theory. 

\subsection{\label{sub:model_function}Model for Staggered Meson Correlation Functions}

We will consider three different types of models for staggered meson two-point correlation functions
in this paper.  As we are employing Bayesian model averaging, further discussed
in Sec.~\ref{sec:model_avg}, we don't have to choose a particular model but rely on the computed
model probabilities to distinguish the most likely models for a given correlation function. 
Within each model type, the number of free parameters in each
specific instance of the model will depend upon the number of oscillating and non-oscillating states
included.

The model we will use for the staggered meson correlation function
in the time domain (\textbf{Model A}) is
\begin{eqnarray}
\label{eq:C_fit}
C(\vec{p},t) & = & c_0 \delta_{\vec{p},0}+ \sum_n \frac{c_n}{2 \left(1 - e^{-E_n N_t}\right) \sinh(E_n)}
\left[ e^{-E_n t} + e^{-E_n(N_t - t)} \right] \nonumber \\
&& + ( -1)^t \sum_j \frac{c^\prime_j}{2 \left(1 - e^{-E^\prime_j N_t}\right) \sinh(E^\prime_j)}
\left[ e^{-E^\prime_j t} + e^{-E^\prime_j (N_t - t)} \right]
\end{eqnarray}
where we have chosen to use a particular ``relativistic'' normalization for the amplitudes.
As is typical for the staggered fermions, there are a set of states labeled by $n$ whose contributions
do not oscillate in time and another set of states labeled by $j$, with different quantum numbers,
that oscillate in time with a factor $(-1)^t$.  The energies $E_n$ and $E_j^\prime$ are understood
to depend implicitly on the spatial momentum $\vec{p}$.  We also allow for the possibility of a
$t$-independent contribution to the correlation function, $c_0$, which is generally not present
for flavor-nonsinglet correlation functions due to translation invariance of the ensemble average.
But, it is the dominant contribution to the flavor-singlet $\sigma$ correlation function and must be
treated carefully in order to extract reliable estimates of model parameters.  Note that the constant
only contributes to the $\vec{p}=0$ correlator, so one method of dealing with this constant is to work
with $\vec{p} \ne 0$ correlators.  Given that we are interested in the energy of the $\sigma$ meson
in the rest frame, $\lim_{\vec{p} \to 0} E_\sigma(\vec{p}) = M_\sigma$, this approach requires a good
understanding of the dispersion relation on the lattice.

To motivate the normalization of amplitudes $c_n$ and $c_j^\prime$ in Eq.~(\ref{eq:C_fit}),
we can perform the discrete cosine transform (DCT-I) of the time-domain
correlation function into the frequency domain analytically
\begin{equation}
\label{eq:C_fit_freq}
\widetilde{C}(\vec{p},k) = c_0 \delta_{\vec{p},0} \delta_{k,0}
+ \frac{1}{N_t} \sum_n \frac{c_n}{\widehat{E}_n^2 + \widehat{\omega}_k^2}
+ \frac{1}{N_t} \sum_j \frac{c^\prime_j}{\widehat{E}^{\prime 2}_j + \widehat{\omega}^{\prime 2}_k}
\end{equation}
where
\begin{equation}
\widehat{E}_n = 2 \sinh \frac{E_n}{2} , \qquad
\widehat{\omega}_k = 2 \sin \frac{2 \pi k}{2 N_t} , \qquad
\widehat{\omega}^\prime_k = 2 \sin\left( \frac{\pi}{2} - \frac{2 \pi k}{2 N_t} \right) .
\end{equation}
Comparing the expression for two different spatial momenta $\vec{p}$, the energies
$E_n$ and $E_j^\prime$ will be different, defining some lattice dispersion relation.
But the amplitudes $c_n$ and $c_j^\prime$ are momentum independent as normalized,
and therefore frame-independent as expected in a Lorentz-invariant theory,
hence a ``relativistic'' normalization.

In our previous work, we considered another method of dealing with the constant $c_0$ which was
to analyze the finite difference correlation function for the $\vec{p}=0$ $\sigma$ meson
\begin{equation}
\Delta_\sigma(t) = C_\sigma(t+1) - C_\sigma(t) .
\end{equation}
In the model, the cancellation of $c_0$ is exact but in our lattice calculation there is inherent
statistical noise contributing to each time slice, so the cancellation is not exact.  In this work,
we propose an improved subtraction scheme for $\vec{p}=0$ correlation functions
\begin{equation}
\overline{C}(t) = C(t) - \frac{1}{N_t} \sum_{t^\prime=0}^{N_t-1} C(t^\prime)
\end{equation}
for states that have a time-independent part, like the $\sigma$ meson.
Given our frequency analysis above,
we can see the subtraction is the zero-frequency component
of the correlation function $\overline{C}(t) = C(t) - \widetilde{C}(0)$.
Furthermore, we know explicitly the functional form of the residual constant
that comes from the integral of the $t$-dependent part of the correlation
function
\begin{equation}
c_0 - \widetilde{C}(0) =
- \frac{1}{N_t} \sum_{n=1}^\infty \frac{c_n}{\widehat{M}_n^2}
- \frac{1}{N_t} \sum_{j=1}^\infty \frac{c^\prime_j}{4 + \widehat{M}^{\prime 2}_j}
\end{equation}
Because some of the fit parameters appear in the residual constant, we will
include that part in the fit and shift the constant (\textbf{Model B})
\begin{eqnarray}
\label{eq:C_sub_fit}
\overline{C}(t) & = & \overline{c}_0
+ \sum_n^{n_\text{max}} \frac{c_n}{2 \left(1 - e^{-M_n N_t}\right) \sinh(M_n)}
\left[ e^{-M_n t} + e^{-M_n(N_t - t)} \right]
- \frac{c_n}{N_t \widehat{M}_n^2} \nonumber \\
&& + \sum_j^{j_\text{max}} \frac{(-1)^t c^\prime_j}{2 \left(1 - e^{-M^\prime_j N_t}\right) \sinh(M^\prime_j)}
\left[ e^{-M^\prime_j t} + e^{-M^\prime_j (N_t - t)} \right]
- \frac{c_j^\prime}{N_t ( 4 + \widehat{M}_j^{\prime 2} ) } \\
\overline{c}_0 & = & - \frac{1}{N_t} \sum_{n=n_\text{max}+1}^\infty \frac{c_n}{\widehat{M}_n^2}
- \frac{1}{N_t} \sum_{j=j_\text{max}+1}^\infty \frac{c^\prime_j}{4 + \widehat{M}^{\prime 2}_j} .
\nonumber
\end{eqnarray}
In counting free parameters, model B will have one more free parameter than model A
and the interpretation of the value of this parameter, $\overline{c}_0$, will depend strongly
on the choice of $n_\text{max}$ and $j_\text{max}$.  In particular, we expect
$\overline{c}_0 \to 0$ within statistical uncertainties as the number of states included
in a particular model instance approaches the limit of available statistics to properly
constrain them.

We will also consider a modification of model B (\textbf{Model C}) where we constrain
$\overline{c}_0 = 0$.  It will have the same number of free parameters as model A
in an instance where they include the same number of states.  In the context of Bayesian
model averaging, we expect that model B will have a higher relative probability than model C
in instances where $\overline{c}_0$ is statistically non-zero.  But with increasing
numbers of states eventually model C
should become more probable, also indicating the limit in which the power
of the available statistics to constrain parameters has been exhausted.

\subsection{\label{sub:dispersion_relation}Staggered Meson Dispersion Relation}

The functional momentum dependence of energies $E_Q(\vec{p})$ extracted
from analysis of two-point correlation functions $C_Q(\vec{p},t)$
is a complicated, non-perturbative problem because Lorentz symmetry is broken
by the lattice discretization so the theory is not invariant under boosts.
Still, Lorentz symmetry
is fully recovered in the continuum limit.  Naively we can expect
\begin{equation}
\label{eq:cont_disp}
a^2 E_Q^2(\vec{p}) = a^2 M_Q^2 + a^2 p^2 + O(a^4 p^4)
\end{equation}
where we explicitly show the lattice spacing $a$ in this dimensionless relation and define the spatial
momentum components $p_i = 2 \pi n_i / (a N_s)$ and
$n_i \in \left\{ - N_s/2 + 1, \cdots, 0, \cdots, N_s/2 \right\}$ and $N_s$ is the number of lattice sites
in the spatial directions.

To improve upon this estimate, one would have to understand the dynamics on the lattice of the eigenstates
corresponding to these energies.  This is a challenging problem since the eigenstates are not simple
single-hadron excitations, in general,  but are more likely strongly-interacting
multi-hadron states.  But, the lowest energy state with given quantum numbers $Q$
may reasonably be expected to behave like
a single-hadron state, particularly if it's energy is well below the nearest multi-hadron threshold.
In this case, we can approximate the dispersion relation with that of a non-interacting boson on the lattice
\cite{Engels:1981ab}
\begin{equation}
\widehat{E}_Q^2 = \widehat{M}_Q^2 + \widehat{p}^2 + O(\widehat{p}^4)
\end{equation}
\begin{equation}
\label{eq:lat_disp}
\widehat{E}_Q = 2 \sinh \frac{a E_Q}{2} , \quad \widehat{M}_Q = 2 \sinh \frac{a M_Q}{2} ,
\quad \widehat{p}_i = 2 \sin \frac{a p_i}{2}
\end{equation}
In the second equation, we have explicitly put in the lattice spacing dependence $a$.
Both lattice dispersion relations correspond to the same continuum relation as $a \to 0$.

In either of these models, Eqs.~(\ref{eq:cont_disp}) or (\ref{eq:lat_disp}), the finite size
of the lattice along spatial directions $N_s$ directly controls the spacing between the discrete
momenta but is not expected to appear explicitly in the finite lattice spacing corrections
$O(a^4 p^4)$ or $O(\widehat{p}^4)$. When we fit our lattice data on two or more volumes
at the same value of the bare coupling and mass, we will parameterize our fits so that the same
lattice corrections are used on all volumes.

\subsection{\label{sub:decay_constants}Staggered Meson Decay Constants}

The normalization in Eq.~(\ref{eq:C_fit}) was chosen such that
$c_n \to \left| \left\langle 0 \left| \mathcal{O} \right| n, \vec{p}=0 \right\rangle
\right|^2$ in the continuum limit with the usual continuum relativistic
normalization.
Following Eq.~(7.5) of \cite{Kilcup:1986dg}
we define the pion decay constant
\begin{equation}
\label{eq:F_pi_5_hat}
\sqrt{2} \widehat{F}_{\pi_5} \left( \widehat{E}_{\pi_5}^2 - \widehat{p}^2 \right) = 2 m_q \frac{1}{\sqrt{N_f}} 
\left\langle 0 \left| P_5 \right| \pi_5(\vec{p}) \right\rangle \quad \implies \quad
\widehat{F}_{\pi_5} = \frac{1}{\sqrt{2}} \ \frac{m_q \sqrt{|c_{\pi_5}|}}{\widehat{E}_{\pi_5}^2 - \widehat{p}^2}
\end{equation}
where $N_f$ in this equation is the number of continuum flavors of a single staggered
fermion, \textit{i.e.} $N_f=4$.  Note we put the hat on the symbol for $\widehat{F}_{\pi_5}$
to indicate the form of the lattice dispersion relation used.  We could have just as easily used
the other form of the lattice dispersion relation, which would lead to a slight different definition
of the decay constant.
Both definitions should converge to the continuum one in the limit of zero lattice spacing.
This definition is slightly different than ones previously used by the LSD Collaboration
for the pion decay constant \cite{LatticeStrongDynamics:2018hun, LatticeStrongDynamics:2021gmp},
but the difference is not statistically significant.

For the isotriplet scalar form factor, there does not seem to be a conventional normalization
\cite{Shi:1999hm, Maltman:2000di} for the decay constant in QCD as it is an unstable resonance.
See review \textit{``Scalar Mesons below 1 GeV''} in \cite{ParticleDataGroup:2020ssz}.
In our $N_f=8$ theory over the range of fermion masses we've studied, the non-singlet
scalar meson appears to be stable, although close in energy to its decay threshold.
We choose to normalize it analogously with the pion decay constant
\begin{equation}
\widehat{F}_{a_{0,1}} =
\frac{1}{\sqrt{2}} \ \frac{m_q \sqrt{|c_{a_{0,1}}|}}{\widehat{E}_{a_{0,1}}^2 - \widehat{p}^2}
\end{equation}
where $c_{a_{0,1}}$ is the residue of the first pole in the frequency domain representation
of the non-singlet scalar two-point correlation function, Eq.~(\ref{eq:C_fit_freq}).

For the isosinglet scalar decay constant, we use the normalization defined
in Eq.~(72) of \cite{LatKMI:2016xxi}
\begin{equation}
\label{eq:scalar_matrix_element}
\widehat{F}_S \ \left( \widehat{E}_\sigma^2 - \widehat{p}^2 \right)
= m_q \left\langle 0 \left| S(0,0) \right| \sigma(\vec{p}) \right\rangle
\end{equation}
where the scalar current is defined as
$S(\vec{x},t) = \sum_{i=1}^{N_f/4} \overline\chi_i(\vec{x},t) \chi_i(\vec{x},t)$.
The two-point correlation function of this scalar current is defined
in Eq.~(\ref{eq:isosinglet}) and, in terms of this correlation function,
the decay constant is defined
\begin{equation}
\label{eq:F_sigma_hat}
\widehat{F}_S =
\frac{m_q \sqrt{|c_\sigma|}}{\widehat{E}_\sigma^2 - \widehat{p}^2} .
\end{equation}
In particular, the normalization used in Eq.~(\ref{eq:isosinglet}) is essential to correctly normalizing the decay constant.

\subsection{\label{sub:finite_volume}Finite Volume Corrections}

In QCD, finite volume corrections to the pion mass and pion decay constant extracted from
a two-point correlation function calculated on a periodic torus of spatial size $L$
can be computed in chiral perturbation theory provided $M_{\pi} L \gg 1$ and
$F_{\pi} L \gg 1$. See Eq.~(6.15) of \cite{Golterman:2009kw}, for example.
In $N_f=8$ over the range of fermion masses for which we have relevant lattice calculations,
chiral perturbation theory is unlikely to be a good effective description
for two reasons: the strong fermion mass dependence of $F_\pi$ and the stable $\sigma$
meson with $M_\sigma \ll 4 \pi F_\pi$.
So it is not expected that finite volume corrections computed
in chiral perturbation theory (ChiPT) will exactly match the numerical calculations.  Still, it seems likely
that whatever low energy effective theory replaces ChiPT will have much the same
structure as these arise from contributions of virtual pion degrees of freedom that probe the finite volume
by wrapping the spatial cycles of the torus, and the pion still is the lightest hadron in the eight-flavor
theory.  There may be additional contributions from $\sigma$-meson degrees of freedom, but they are expected
to be sub-leading due to the somewhat heavier mass.

We will follow the approach used in \cite{LatticeStrongDynamics:2021gmp} and use ChiPT-inspired
forms to model our finite volume corrections
\begin{eqnarray}
\label{eq:M_X_FV}
M_Q(L) & = & M_Q(\infty) \left[
1 + \alpha_Q \frac{M_\pi^2}{(4 \pi F_\pi)^2}
\sum_{n=1}^\infty \frac{4 \ \kappa(n)}{\sqrt{n} \ M_\pi L}
K_1(\sqrt{n} \ M_\pi L)
\right] \\
\label{eq:F_X_FV}
F_Q(L) & = & F_Q(\infty) \left[
1 + \beta_Q \frac{M_\pi^2}{(4 \pi F_\pi)^2}
\sum_{n=1}^\infty \frac{4 \ \kappa(n)}{\sqrt{n} \ M_\pi L}
K_1(\sqrt{n} \ M_\pi L)
\right] .
\end{eqnarray}
The function $\kappa(n)$ counts the number of lattice vectors $\vec{n}$ with integer-valued
components of length $\sqrt{n}$, see Tab.~\ref{tab:kappa}.  In QCD,
it is common to expand the sum over modified Bessel functions $K_1$, assuming $M_\pi L \gg 1$
and keep only the leading term, particularly if $M_\pi L \gtrsim 4$ in all, leading to
\begin{equation}
\sum_{n=1}^\infty \frac{4 \ k(n)}{\sqrt{n} \ M_\pi L}
K_1(\sqrt{n} \ M_\pi L) \approx \frac{12 \sqrt{2 \pi}}{(M_\pi L)^{3/2}} e^{- M_\pi L}
\end{equation}
In an earlier paper \cite{LatticeStrongDynamics:2021gmp}, we also used this approximation
for the finite volume extrapolation of $M_\pi$ and $F_\pi$.  We did not observe any significant change
in the result if we included more terms in the expansion.  In this analysis, we will be conservative
and not expand the modified Bessel functions and truncate the sum only after the first eight terms
(up to $n=8$) although we expect it will not make a significant difference relative to keeping just
the leading term.

\begin{table}[ht]
\centering
\addtolength{\tabcolsep}{3 pt}   
\begin{tabular}{c|c|c|c}
$n$ & $\vec{n}$ & $|\vec{n}|$ & $\kappa({n})$ \\
\hline
0 & (0,0,0) & 0 & 1 \\
1 & (1,0,0) & 1 & 6 \\
2 & (1,1,0) & $\sqrt{2}$ & 12 \\
3 & (1,1,1) & $\sqrt{3}$ & 8 \\
4 & (2,0,0) & 2 & 6 \\
5 & (2,1,0) & $\sqrt{5}$ & 24 \\
6 & (2,1,1) & $\sqrt{6}$ & 24 \\
7 & --- & --- & 0 \\
8 & (2,2,0) & $\sqrt{8}$ & 12 \\
\hline
9 & (2,2,1) & 3 & 24 \\
$9^\prime$ & (3,0,0) & 3 & 6 \\
\end{tabular}
\caption{\label{tab:kappa} The number of lattice vectors $\vec{n}$
with integer-valued components of length $\sqrt{n}$.  Note there are no vectors
of length $\sqrt{7}$ and, starting at length 3, there may be multiple inequivalent
sets of vectors under the cubic group.}
\end{table}

Since the infinite volume extrapolation described in this section implicitly assumes
that the pion is a pseudo-Nambu-Goldstone boson, one should use caution when modeling
the extrapolated data provided later in this paper, particularly if one wants to explore
other finite volume corrections, \text{e.g.}\ due to a light isosinglet scalar.
If one assumes that the massless limit of the theory approaches a conformal fixed point,
possible finite volume corrections were discussed in \cite{Appelquist:2011dp}.
In either case, one should use the finite volume data provided in the supplementary materials
\cite{LatticeStrongDynamics:2023zenodo}
when performing further analysis.

\subsection{\label{sub:GMOR_ratio}The GMOR Relation and Near-Conformality}

As a guide to constructing low energy effective descriptions
$N_f=8$ theory, it would be useful to characterize the extent to which
one or a few light states dominates the low-energy dynamics.
An important phenomenological tool for characterizing the degree to which
the dynamics of the Nambu--Goldstone pions dominates low-energy phenomena
in QCD was first described by Gell-Mann, Oakes and Renner (GMOR)
\cite{Gell-Mann:1968hlm}.  In their original derivation, they \textit{a priori}
assumed pion-pole dominance and derived the GMOR relation as
a consequence.  Our derivation will not initally assume pole-dominance but
start with the integral of the axial Ward--Takehashi identity.
In our notation, this can be written
\begin{equation}
\label{eq:pion_susceptibility}
\sum_{t=0}^{N_t-1} C_{\pi_5}(\vec{0},t) = \frac{1}{m_q}
\text{Tr}_{\text{color}} \left[ G_F(\vec{0},0;\vec{0},0) \right]
\end{equation}
which is an exact spectral identity on each gauge configuration,
not just in the ensemble average.  In the chiral limit $m_q \to 0$
of a theory with spontaneous chiral symmetry breaking, the trace
on the right hand side will approach a constant following the
Banks--Casher relation \cite{Banks:1979yr, Leutwyler:1992yt} and
the integrated pion correlation function will diverge due to the massless
Nambu--Goldstone pion.  Using Eqs.~(\ref{eq:C_fit_freq}) and (\ref{eq:F_pi_5_hat})
we can identify the rate of this divergence with parameters in our fit functions
\begin{equation}
\sum_{t=0}^{N_t-1} C_{\pi_5}(\vec{0},t) \to \frac{c_{\pi_5}}{\widehat{M}_{\pi_5}^2}
= 2 \frac{\widehat{F}_{\pi_5}^2 \widehat{M}_{\pi_5}^2}{m_q^2} \quad \text{as} \quad m_q \to 0 .
\end{equation}
Using the normalization of the isosinglet scalar current
in Eq.~(\ref{eq:scalar_matrix_element})
leads to a generalization of the GMOR relation for general $N_f$
\begin{equation}
\label{eq:generalized_GMOR}
m_q \left\langle S \right\rangle
= m_q \frac{N_f}{4} \left\langle \overline\chi \chi \right\rangle \ge
\frac{N_f}{2} \widehat{F}_{\pi_5}^2 \widehat{M}_{\pi_5}^2 .
\end{equation}
Now, if we assume spontaneous symmetry breaking and pion pole-dominance,
the inequality becomes an equality in the limit $m_q \to 0$
and $\left\langle S \right\rangle$ approaches a well-defined low energy constant,
which is the usual GMOR relation.

Patella \cite{Patella:2011jr} has noted that Eq.~(\ref{eq:generalized_GMOR})
should also be true in a mass-deformed CFT with a large mass anomalous dimension
$( 1 < \gamma^* < 2)$
due to a large contribution
to the pion correlation function generated by the running of the mass.
They propose examining the GMOR ratio
\begin{equation}
\label{eq:GMOR_ratio}
R_G(m_q) \equiv \frac{m_q \left\langle \overline\chi \chi \right\rangle}
{2 \widehat{F}_{\pi_5}^2 \widehat{M}_{\pi_5}^2}
= \left\{\begin{array}{ll}
1 \ , & \text{(near-conformal)} \\
1 < R_G(0) < \infty \ , & (\text{CFT}, 1 < \gamma^* < 2) \\
\infty \ , & (\text{CFT}, 0 < \gamma^* < 1)
\end{array} \right. \quad \text{as}
\quad m_q \to 0
\end{equation}
for an indication of whether the theory is near-conformal or conformal
in the chiral limit.  In a near-conformal scenario, it is not clear
at what fermion mass $m_q$ one would expect to see the transition from the
approximately hyperscaling regime where $R_G(m_q) > 1$ to the spontaneously
broken regime where $R_G(m_q) \to 1$ as $m_q \to 0$.
Just observing $R_G(m_q) > 1$ at some finite fermion mass
is not sufficient to establish IR conformality. In particular, one must follow
the correct order of limits: volume to infinity, lattice spacing to zero,
and then fermion mass to zero.

\section{\label{sec:model_avg}Bayesian Model Averaging}

\subsection{\label{sub:model_avg_setup}General Setup}

One of the challenges observed in our previous analysis of the light meson spectrum
in the $N_f=8$ theory
\cite{LatticeStrongDynamics:2018hun} were large systematic errors due to fit parameters
varying significantly over a range of different fits while $\chi^2 / \text{dof}$ did not.
We define $\log p(D|M)$ by the usual
chi-squared prescription
\begin{equation}
\label{eq:chisq}
\log p(D|M) \propto - \frac{1}{2} \sum_{t,t^\prime \in T_1}
\left( C(t) - f_M(t) \right) \Sigma_{t t^\prime}^{-1} \left( C(t^\prime) - f_M(t^\prime) \right)
\end{equation}
where $C(t)$ is correlation function computed from our lattice ensemble $D$, $f_M(t)$ is the function
for model $M$ to be fitted by minimizing the log-likelihood, $T_1$ is the subset of times selected
for fitting and $\Sigma_{t t^\prime}$ is the covariance of the correlation function $C(t)$ on the subset $T_1$.
Assuming all the quantities are properly estimated from the ensemble, the log-likelihood is expected
to sample the chi-squared distribution for degrees of freedom equal to the number of times in $T_1$
minus the number of free parameters in $M$.

Subsequent to our earlier analysis, Jay and Neil proposed \cite{Jay:2020jkz} a Bayesian model
averaging analysis framework which estimates $\log p(M|D)$, the probability that a model $M$
is a good representation of the data selection $D$.  One suggested estimator
of the model probability
is based on the Akaike Information Criterion (AIC), provided nuisance model parameters are assigned
to account for data subsets not included in the fit.  For example, let $M$ be a model
with $N_M$ free parameters and the maximal data set has $N_T$ times available to be fit.
If we perform the fit only on a subset of times $T_1$ of size $N_1$ then the number
of data points not included $N_0 = N_T - N_1$ must be assigned nuisance parameters.
Thus, for the AIC, the number of relevant
parameters is $N_M + N_0$ and the model probability \cite{Neil:2023pgt}
is
\begin{equation}
\log p(M|D) \propto \log p(D|M) - (N_M + N_0)
\end{equation}
After the model probability has been estimated for the full set of models $\{M\}$ to be considered for the
analysis, we normalize this set of probabilities: $\sum_{\{M\}} p(M|D) = 1$.  In App.~\ref{sec:prob_norm},
we provide some details how we perform this sum accurately given the potential for widely varying values
of $\log p(M|D)$.

With an reasonable estimate of the model probability, it seems straightforward to construct expectation values
and variances of model parameters over the set of possible models considered.  For example, the expected
value of a model parameter is
\begin{eqnarray}
E(a) & = &  \frac{1}{\Sigma_1}
\sum_{\{M|a \in M\}} a_M \ p(M|D) \ \Theta\left[ p(M|D) - p_\text{cut} \right]
\nonumber \\*
\Sigma_1 & = & \sum_{\{M|a \in M\}} p(M|D) \ \Theta\left[ p(M|D) - p_\text{cut} \right]
\end{eqnarray}
where, to make sure the notation is clear, we compute a weighted average over only the subset of models
that contain the parameter, $\{M | a \in M\}$ and further consider only models where the model weight
is greater than some pre-determined minimum $p_\text{cut}$, as enforced by the Heavyside function $\Theta$.

The variance of the model-averaged expectation value has two contributions.  The first, and usually dominant,
contribution is the weighted average over models of the square of the error estimate $\sigma_{a,M}$
for the parameter $a_M$ in a given model $M$
\begin{equation}
E(\sigma^2_a) = \frac{1}{\Sigma_1}
\sum_{\{M|a \in M\}} \sigma^2_{a,M} \ p(M|D) \ \Theta\left[ p(M|D) - p_\text{cut} \right]
\end{equation}
The second, usually sub-dominant, contribution is the weighted variance of the model estimates
of parameters $a_M$, relative to the model-averaged expectation $E(a)$
\begin{eqnarray}
\text{Var}(a) & = & \frac{\Sigma_1}{\Sigma_1^2 - \Sigma_2}
\sum_{\{M|a \in M\}} \left( a_M - E(a) \right)^2 \ p(M|D) \ \Theta\left[ p(M|D) - p_\text{cut} \right]
\nonumber \\*
\Sigma_2 & = & \sum_{\{M|a \in M\}} p(M|D)^2 \ \Theta\left[ p(M|D) - p_\text{cut} \right]
\end{eqnarray}
The final error estimate for the model average of a parameter is to add the two contributions
in quadrature
\begin{equation}
\sigma_a = \sqrt{ E(\sigma^2_a) + \text{Var}(a) }
\end{equation}

Now we can discuss the motivation behind the probability cut $p_\text{cut}$.  In our experience,
the model-averaged $E(a)$ tend to be dominated by a few choices whose $p(M|D) \sim O(1)$. It seems
reasonable to expect that $E(\sigma_a^2)$ should be similarly dominated by choices
whose $p(M|D) \sim O(1)$ and not $p(M|D) \sim O(p_\text{cut})$.  However, we have observed cases
of overfitting for certain models where as the data selection changes such that $p(M|D)$ decreases,
$\sigma_{a,M}^2$ increases at a faster rate, leading to those very unlikely model choices
to dominate the model average of the squared error $E(\sigma_a^2)$. $p_\text{cut}$ can be adjusted
to minimize the impact of this scenario.

To understand how this can happen, we recall that uncertainty of a two-point meson correlation function
grows exponentially in Euclidean time \cite{Lepage:1989hd}
\begin{equation}
\text{Var}\left[ C_Q(\vec{p},t) \right]
\sim \exp\left[ 2 \left( E_Q(\vec{p}) - M_{\pi_5} \right) t \right]
\end{equation}
Now, for a given model function $M$ with its fixed number of exponential terms, there is a certain
$t_\text{min}$ for which $ - \chi^2/2 \sim (N_M - t_\text{max} + t_\text{min} - 1)/2$,
indicating a good fit using the usual chi-squared criteria $\chi^2/\text{dof} \sim 1$.
For fits on the interval $[t,t_\text{max}],
t < t_\text{min}$, there will be no good fits according to chi-squared, whereas
for fits on the interval
$[t,t_\text{max}], t > t_\text{min}$, $-\chi^2/2$ will approximately increase by $(t-t_\text{min})/2$
indicating continued goodness-of-fit.  However, as the minimum $t$ increases in a given fit, the number
of times not included in the fit also increases: $\Delta N_0 = t - t_\text{min}$.  The net effect
of increasing $t > t_\text{min}$ is to decrease $p(M|D) \propto \exp( -(t-t_\text{min})/2)$.
If $E_Q(\vec{p}) - M_{\pi_5} > 1/4$ we expect that the uncertainties
in model parameters will grow faster
than the model probability decreases as $t > t_\text{min}$.  Based on these
considerations, we have found $p_\text{cut} = 10^{-3}$ is a reasonable choice
for this analysis and we adopt it throughout.  While this analysis was nearing
completion, an alternate approach to dealing with these challenges was proposed
\cite{Neil:2022joj}.  It would be interesting to compare these two approaches
in future analyses.

In our analysis of $I=2$ $\pi_5 \pi_5$ scattering \cite{LatticeStrongDynamics:2021gmp},
we implemented Bayesian model averaging.  As we had hoped, the systematic uncertainties for $\pi_5$-related
observables were greatly reduced in that paper relative to earlier paper \cite{LatticeStrongDynamics:2018hun}.
Also, the problem with uncertainties increasing for $t > t_\text{min}$ was not apparent because we were considering
primarily $\pi_5$-related observables.  We expect this will not be the case for $\sigma$ and $a_{0,1}$-related
observables.

\subsection{\label{sub:shrinkage}Shrinkage Estimator of Covariance}

Suppose one wants to estimate from a multivariate sample a particular element
of the  covariance matrix,
then one usually uses the standard unbiased sample estimator
\begin{equation}
\Sigma_{ij} = \frac{1}{N-1} \sum_{n=1}^N (x_i^{(n)} - \bar{x}_i) (x_j^{(n)} - \bar{x}_j)
\end{equation}
which is derived from the maximum likelihood estimate (MLE) of covariance
of a multivariate Gaussian distribution. By the central limit theorem, as $N \to \infty$
the standard estimator approaches the MLE for any distribution.  To estimate a full
$\mathbb{R}^{K \times K}$ covariance matrix, there are $K(K+1)/2$ independent matrix elements
that must be estimated, requiring $N$ independent samples for each one.  Furthermore, accurate
estimation of the covariance is crucial when using the chi-squared prescription in
Eq.~(\ref{eq:chisq}) since the inverse of the covariance matrix is used and the consequence
of poorly-estimated small eigenmodes is amplified.  Empirically, it has been found that
approximately $50 K(K+1)/2$ samples are needed in lattice QCD calculations for the standard
estimator to be sufficiently accurate for chi-squared fitting
\cite{Michael:1993yj, Simone:2017} .

If you only care about this particular matrix element, or perhaps one more,
then this is the optimal estimator to use.  However, if you want to simultaneously
estimate three or more elements of the covariance matrix, Stein \cite{Stein:1956} proved
that this was not the optimal estimator in the sense
of minimizing the combined mean square error,
\textit{i.e.}\ $\sum_{ij} (\Sigma_{ij} - \Sigma_{ij}^*)^2$
where $\Sigma_{ij}^*$ is the true but unknown covariance.  This was so counterintuitive
at the time, it was called \textsl{Stein's paradox}.

For our purposes, Stein's improved estimator will take the form of the linear shrinkage estimator
of covariance
\begin{equation}
\sigma_{ij}(\lambda) = \lambda \Sigma_{ii} \delta_{ij} + (1 - \lambda) \Sigma_{ij} ,
\quad \lambda \in [0,1]
\end{equation}
and for a given sample ensemble, there exists some optimal $\lambda^*$ that minimizes
the MSE and $\lambda^* \to 0$ as $N \to \infty$.  Since we don't know the true covariance
$\Sigma^*$ we must estimate the optimal value.  Based on work by Ledoit and Wolf \cite{Ledoit:2004},
Sch{\"a}fer and Strimmer \cite{Schafer:2005} gave a fairly straightforward estimator for optimal
value of $\lambda$
\begin{equation}
\hat\lambda^* = \frac{\sum_{i \le j} \widehat{\Var}(\Sigma_{ij})}{\sum_{i \ne j} \Sigma_{ij}^2}
\end{equation}
In App.~\ref{sec:VarCov}, we show a one-pass
algorithm to compute the sample estimate of $\widehat{\Var}(\Sigma_{ij})$.

The shrinkage estimator of covariance has been suggested for use in lattice quantum field theory (LQFT)
applications for some time \cite{Lichtl:2007, Cohen:2010}.  Only recently has the shrinkage estimator
been actually employed for use in published lattice QCD analyses
\cite{Rinaldi:2019thf, NPLQCD:2020ozd, Shanahan:2020zxr,
FermilabLattice:2022gku}. Recent work by Ledoit and Wolff
\cite{Ledoit:2018, Ledoit:2020} have proposed an improved non-linear shrinkage estimator.
Burda and Jarosz \cite{Burda:2021} have also developed an improved shrinkage estimator
and have developed an open-source Python library called \texttt{shrinkage} to assist
in calculations.  In this analysis, we have conservatively chosen to use linear shrinkage
rather than one of the newer alternatives.

\section{\label{sec:detailed_example}Detailed Example of Model-Averaging
Analysis on a Single Ensemble}

We will discuss in detail our analysis of the $96^3 \times 192$, $\beta=4.8$, $m=0.00125$ ensemble which is
the larger volume companion to the $64^3 \times 128$ ensembles discussed in our previous work
\cite{LatticeStrongDynamics:2018hun, LatticeStrongDynamics:2021gmp}.  It will also serve as a detailed example
of how we implemented our model-averaging analysis.

\subsection{Data Selection}

In order to compare models fit to different data subsets, we need to first identify the maximal data set $T$
which could be considered for any model.  Although our staggered meson two-point correlation function data
is computed from $t=0$ to $t=N_t-1=191$, the data are first symmetrized: $(C(t) + C(N_t - t))/2 \to C(t)$
and now the largest possible data set is from $t=0$ to $t=N_t/2=96$. As already mentioned \cite{Lepage:1989hd},
the signal-to-noise decreases exponentially at large times, so for most correlators, particularly at non-zero
momenta $\vec{p} \ne 0$, there is insufficient signal to reasonably include those data points in the analysis
particularly since this will exacerbate the problem of reliable covariance estimation. We will not use data
for $t=0,1$ given the difficulties of interpreting a staggered correlation function separated by one unit in time
in terms of a transfer matrix \cite{Kilcup:1986dg}.
We compute the jackknife ratio
$C_Q(\vec{p},t) / C_Q(\vec{p},1)$ and choose a minimum value for this quantity for each state $Q$ where there is still
good signal-to-noise for all $\vec{p}$.  This defines $t_\text{max}$ for each $Q$ and
$\vec{p}$.

\begin{figure}
    \centering
    \includegraphics[width=0.48\textwidth]{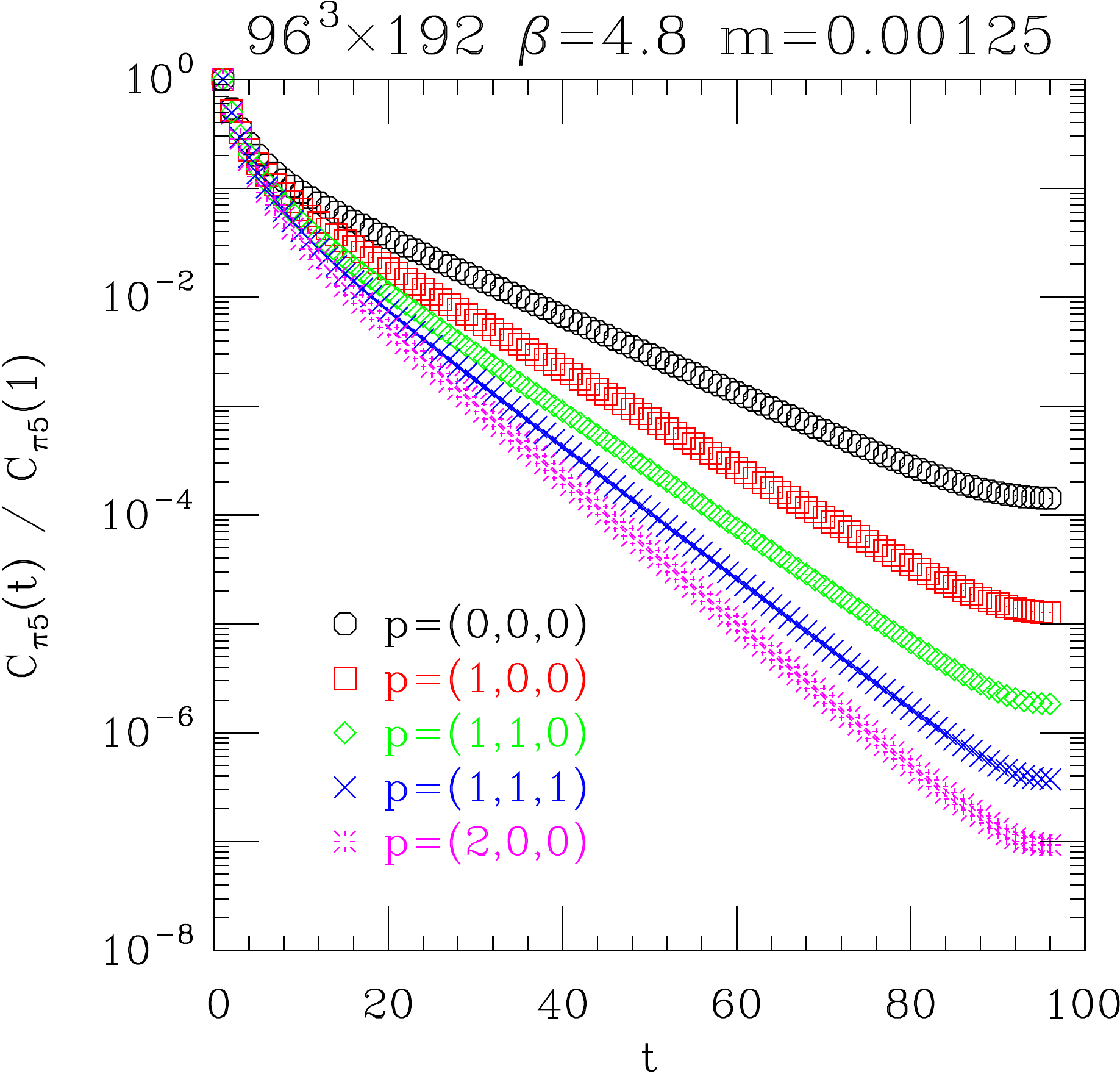}
    \includegraphics[width=0.48\textwidth]{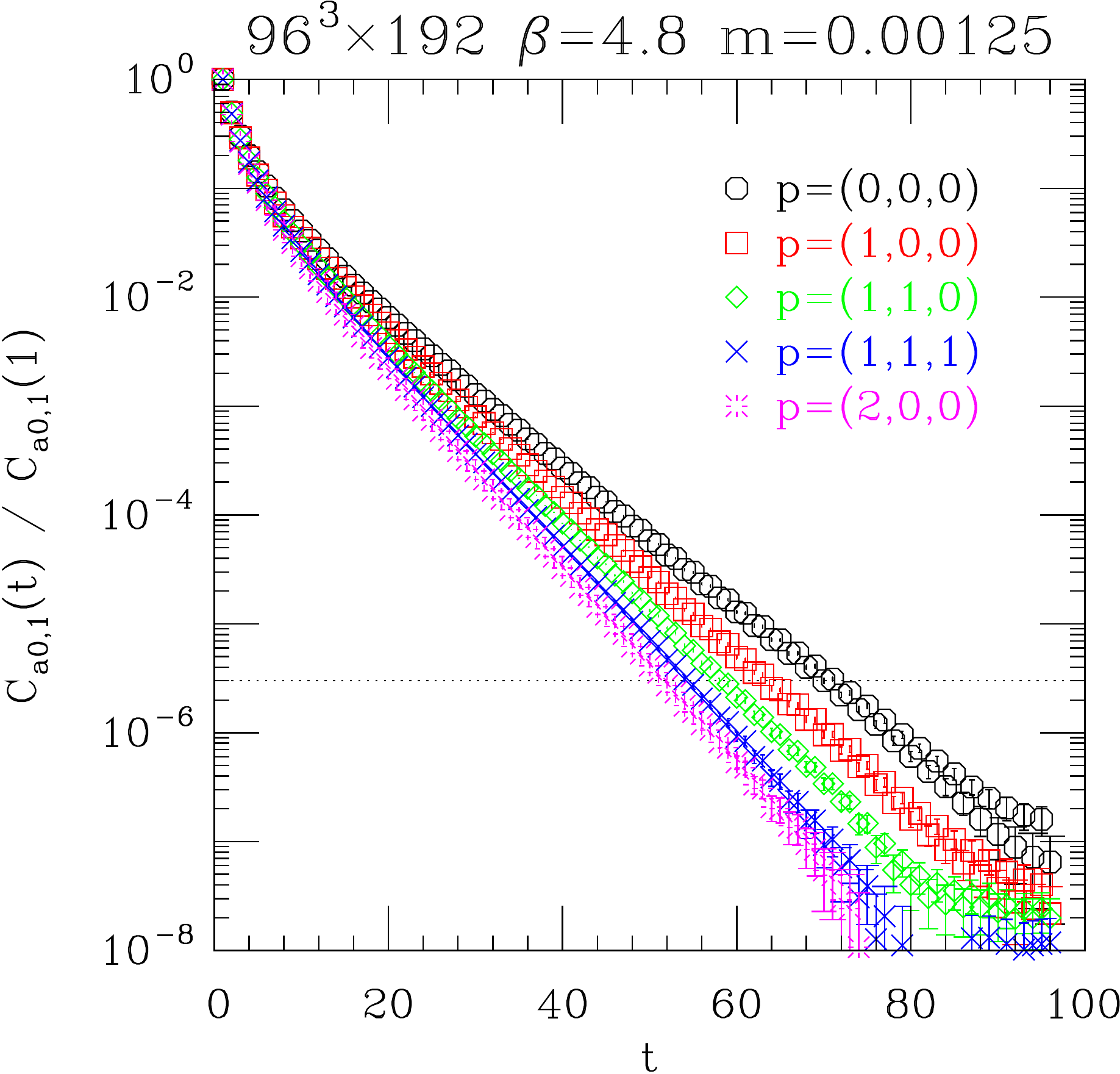}
    \caption{\label{fig:corrPC_rel}Two-point correlation functions for $\pi_5$ and $a_{0,1}$ mesons.
    In the right panel, data points below the horizontal line at $3 \times 10^{-6}$ were not included in any fits.}
\end{figure}

Fig.~\ref{fig:corrPC_rel} shows examples of our procedure.  On the left, for the $\pi_5$ meson, we see good signal
for all momenta to the middle of the lattice and we also see nice straight lines on the log plot indicating clear signal
of a single decaying exponential.  On the right, for the $a_{0,1}$ meson, the situation is somewhat different.  There does
seem to be pretty good signal to the middle of the lattice, but the nature of the signal changes at large times, with an
apparent change of slope and an oscillating signal becoming dominant.  We use a rough model
to guide our choice of where to draw a horizontal line based on the dispersion relation
$E_Q^2 = M_Q^2 + p^2$ and assuming that a single exponential dominates
the correlation function  at times $t_c$ where it crosses the line
\begin{equation}
\label{eq:tc_model}
\frac{C_Q(\vec{p},t_c)}{C_Q(\vec{p},1)} = e^{-\sqrt{M_Q^2 + p^2}(t_c-1)} = \text{const} \quad
\implies \quad t_c(\vec{p}) \propto \frac{1}{\sqrt{M_{\text{eff}}^2 + p^2}}
\end{equation}
We compare the computed values to this model and we see good agreement along
the shown cut line.  If we lower the cut line, the observed values deviate
from the prediction, particularly for $p^2=4$, so we conservatively set
the cut line at $3 \times 10^{-6}$.  Note, this model will not work well as
$t_c \to N_t/2$ since it does not include the additional contribution due
to periodic boundary conditions which becomes important in that region.
A modified expression involving hyperbolic cosines can be derived but we didn't
need it here.

\begin{figure}
    \centering
    \includegraphics[width=0.32\textwidth]{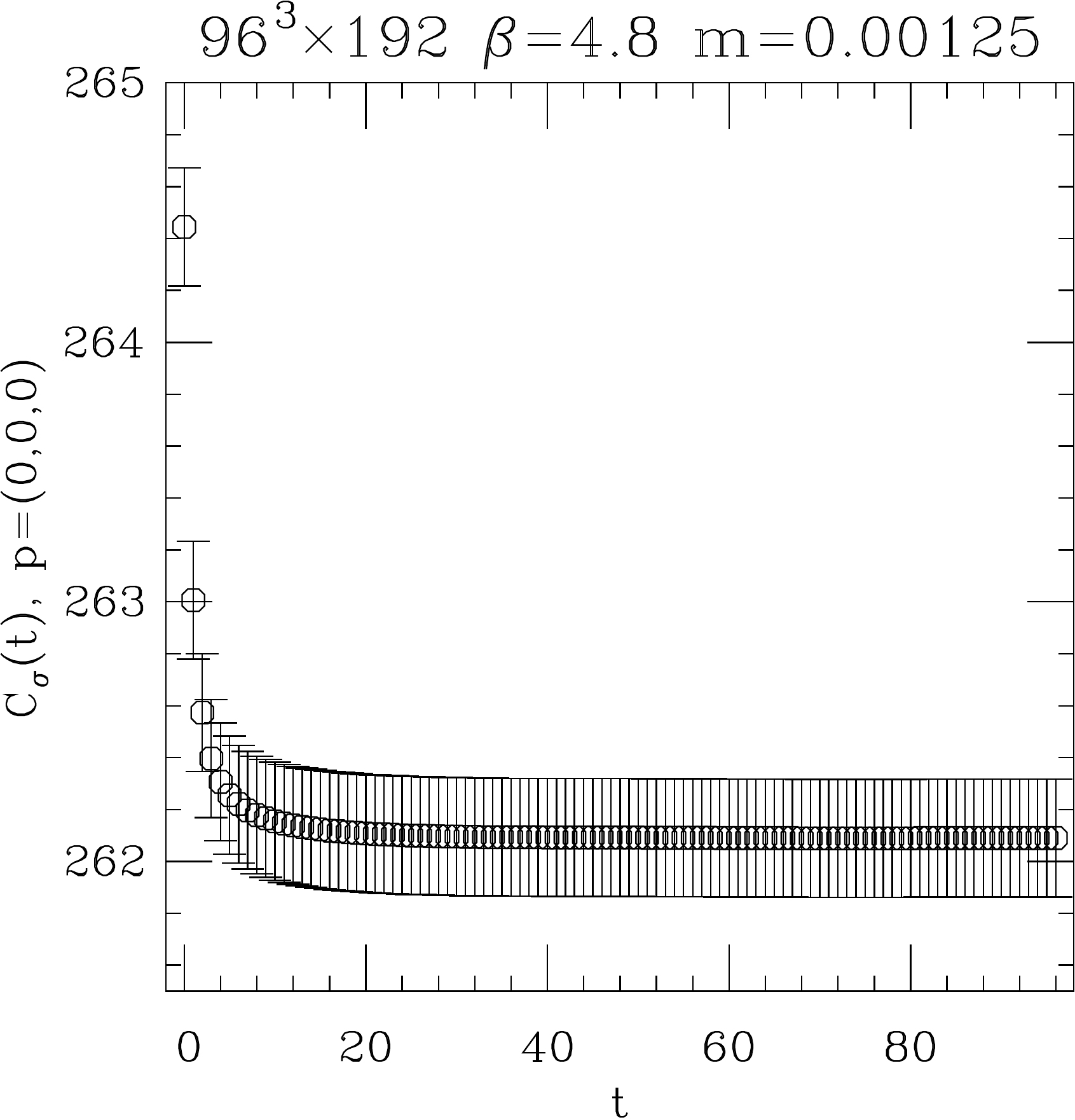}
    \includegraphics[width=0.32\textwidth]{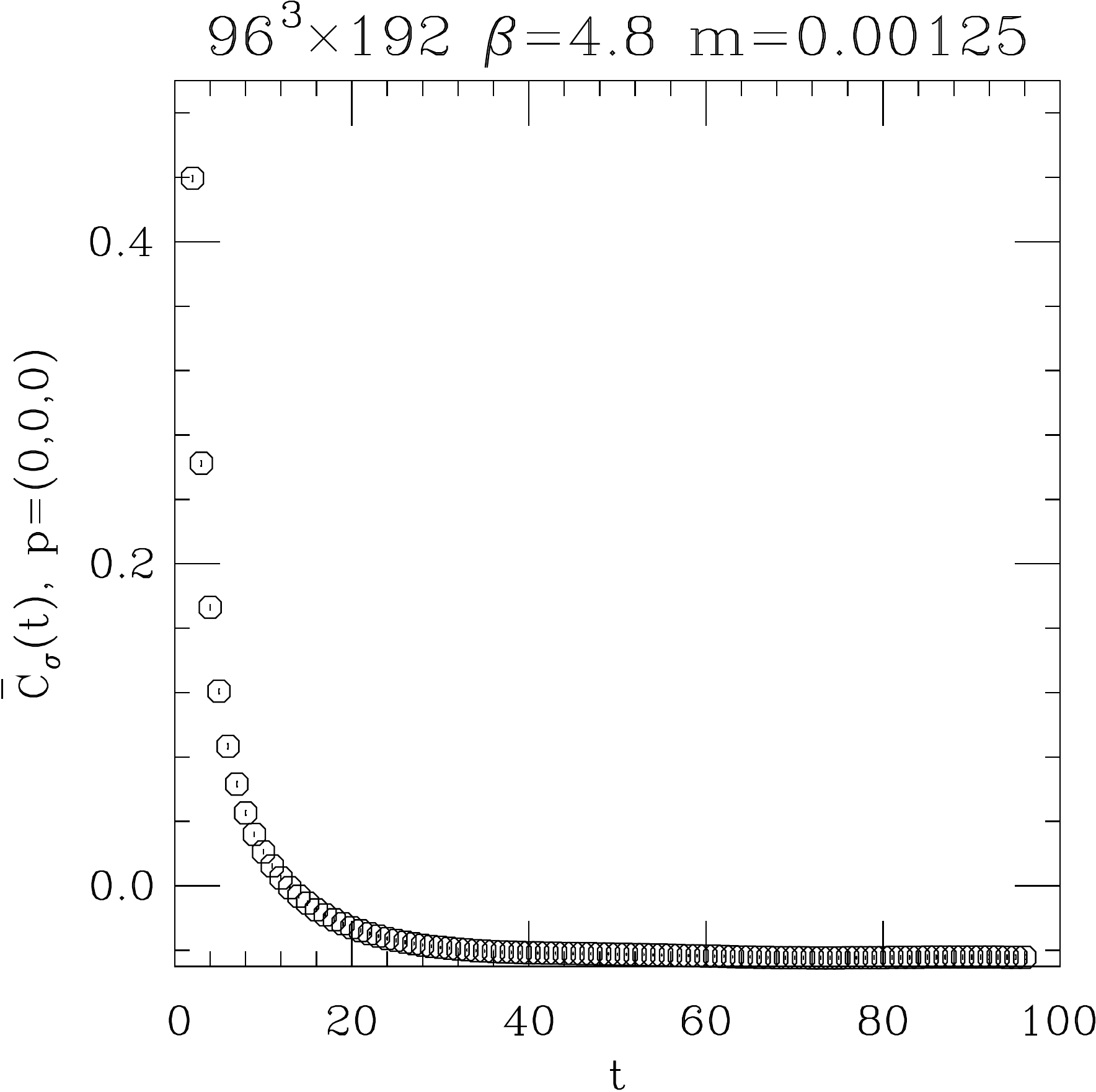}
    \includegraphics[width=0.32\textwidth]{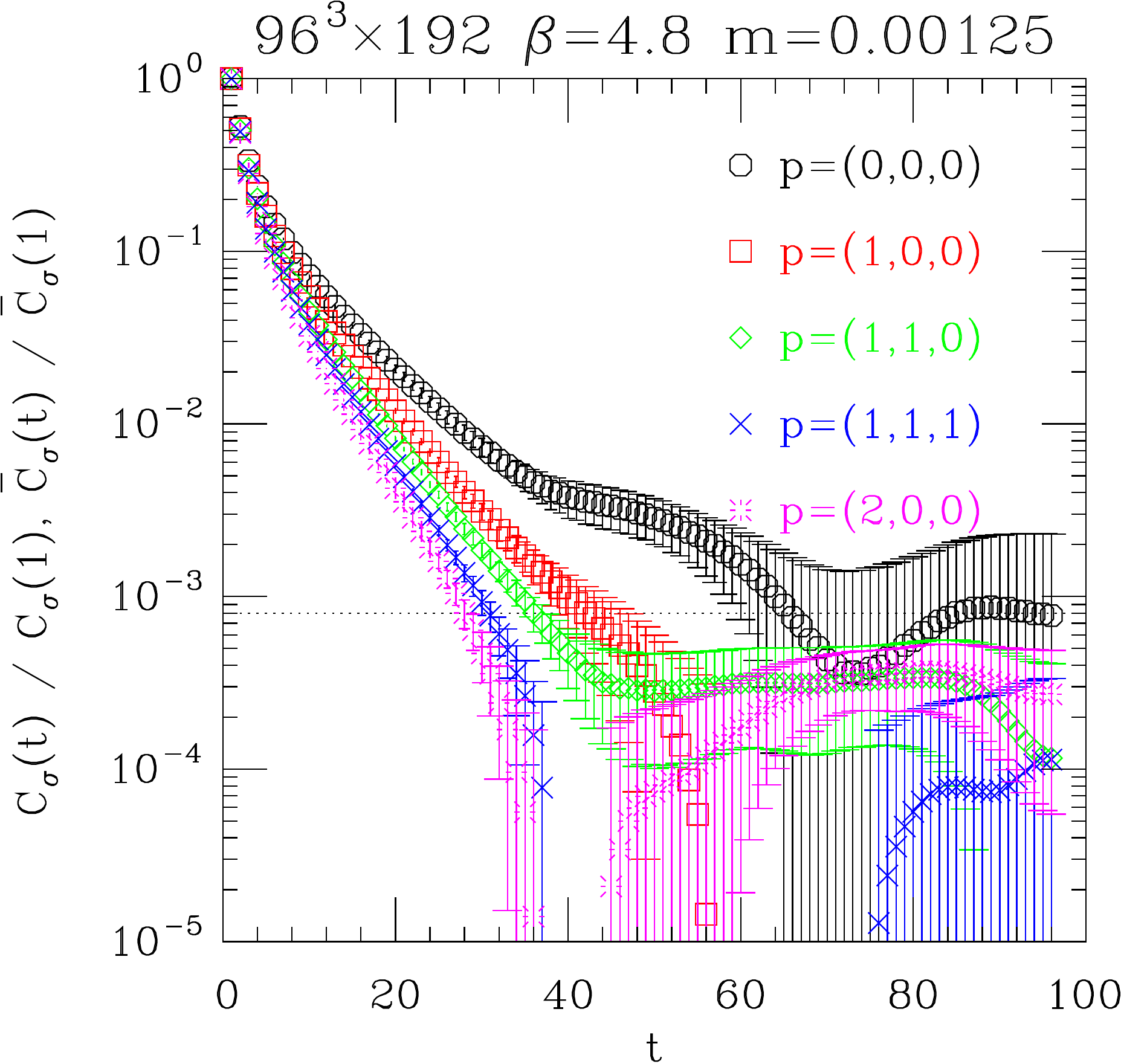}
    \caption{\label{fig:corrS}Two-point correlation functions for $\sigma$ meson.
    Left panel shows the unsubtracted $\vec{p}=0$ correlator. Center panel shows
    $\vec{p}=0$ subtracted correlator.  Right panel data points below horizontal
    line for $\vec{p} \ne 0$ not included in fits.  The data from the central
    panel is included on the right by shifting upwards by a sufficiently large
    constant $\overline{C}(\vec{p}=0,t) + 0.045$ so that the result is positive
    and can be displayed on a log plot. The shifted data cannot be used in the data
    selection analysis.
    }
\end{figure}

The situation for the $\sigma$ meson correlator is more complicated. In Fig.~\ref{fig:corrS} on the left is the
unsubtracted correlator $C_\sigma(\vec{p}=0,t)$.  It should be clear that just subtracting some constant value around
$c_0 = 262.08\cdots$ in an uncorrelated way, following Eq.~(\ref{eq:C_fit}), would be unsatisfactory
because the signal-to-noise would fall below one in a few time units.  The center panel shows
$\overline{C}_\sigma(\vec{p}=0,t)$ and, following Eq.~(\ref{eq:C_sub_fit}), the previously large positive constant
has been replaced with a three orders of magnitude smaller negative constant and greatly enhanced signal-to-noise.
However, we still need to figure out at what time $t_c$ the signal-to-noise of the exponentially decaying part
of the correlator falls below an acceptable level.  We cannot judge this from the central panel since the large
time behavior is dominated by the integral of the correlation function.  Instead, we compute the ratio
$C_\sigma(\vec{p},t) / C_\sigma(\vec{p},1)$ for $\vec{p} \ne 0$ and rely on our crude model Eq.~(\ref{eq:tc_model})
to extrapolate to $\vec{p}=0$, shown in right panel.  The results in the data selection procedure are summarized
in Tab.~\ref{tab:summ_t_range}.

\begin{table}[h]
\centering
\addtolength{\tabcolsep}{3 pt}   
\begin{tabular}{c|c|c|c|c|c}
& $\vec{p}=(0,0,0)$ & $\vec{p}=(1,0,0)$ & $\vec{p}=(1,1,0)$ & $\vec{p}=(1,1,1)$ & $\vec{p}=(2,0,0)$ \\
\hline
$\pi_5$ & [2,96] & [2,96] & [2,96] & [2,96] & [2,96] \\
$a_{0,1}$ & [2,70] & [2,63] & [2,58] & [2,53] & [2,50] \\
$\sigma$ & [2,52] & [2,41] & [2,34] & [2,30] & [2,27] \\
\end{tabular}
\caption{\label{tab:summ_t_range} Summary of maximum allowed time ranges for fitting in model averaging procedure
for the $96^3 \times 192$, $\beta = 4.8$, $m=0.00125$ ensemble.}
\end{table}

\subsection{Model Averaging}

\begin{figure}
\centering
\includegraphics[width=0.32\textwidth]{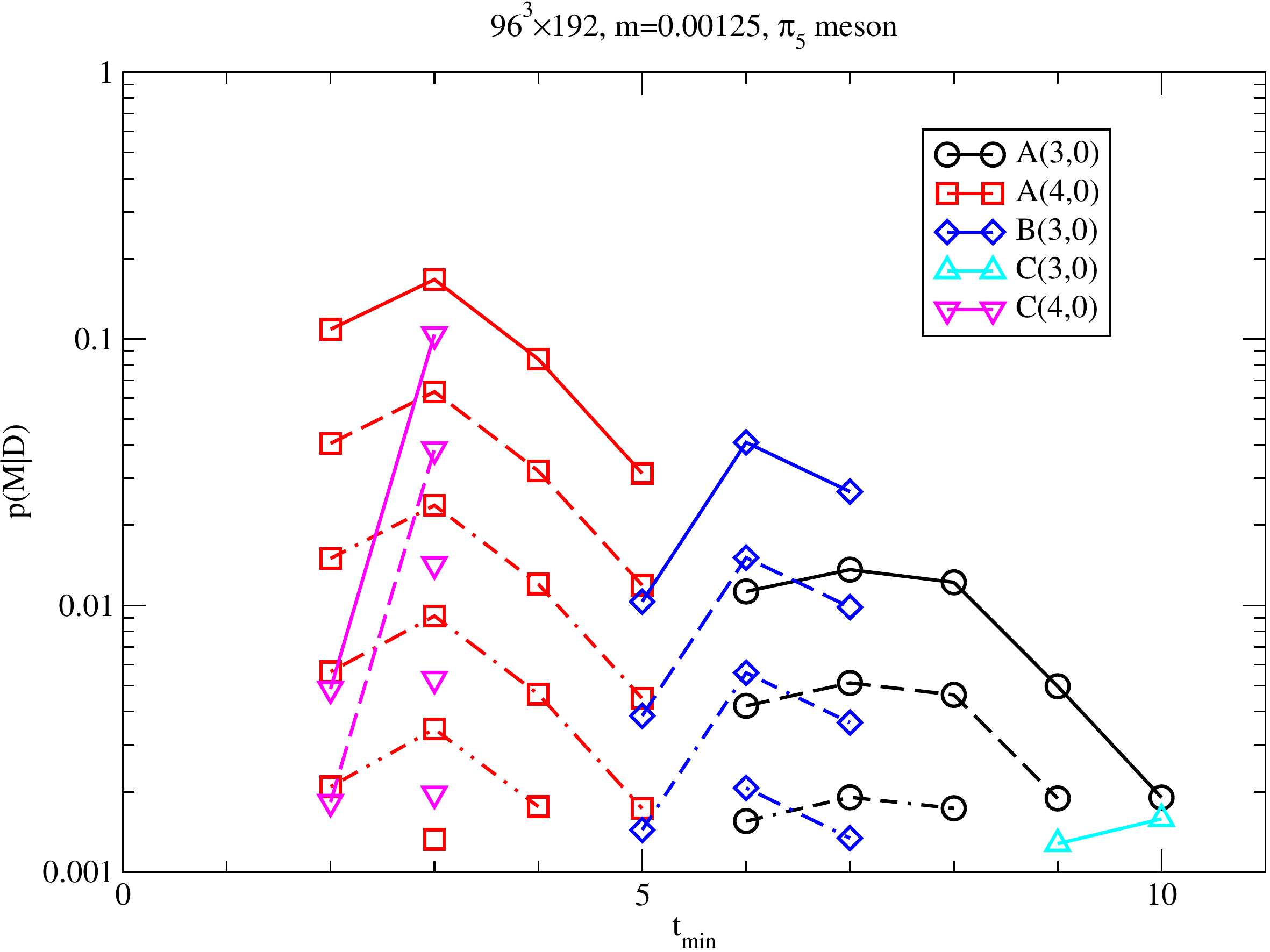}
\includegraphics[width=0.32\textwidth]{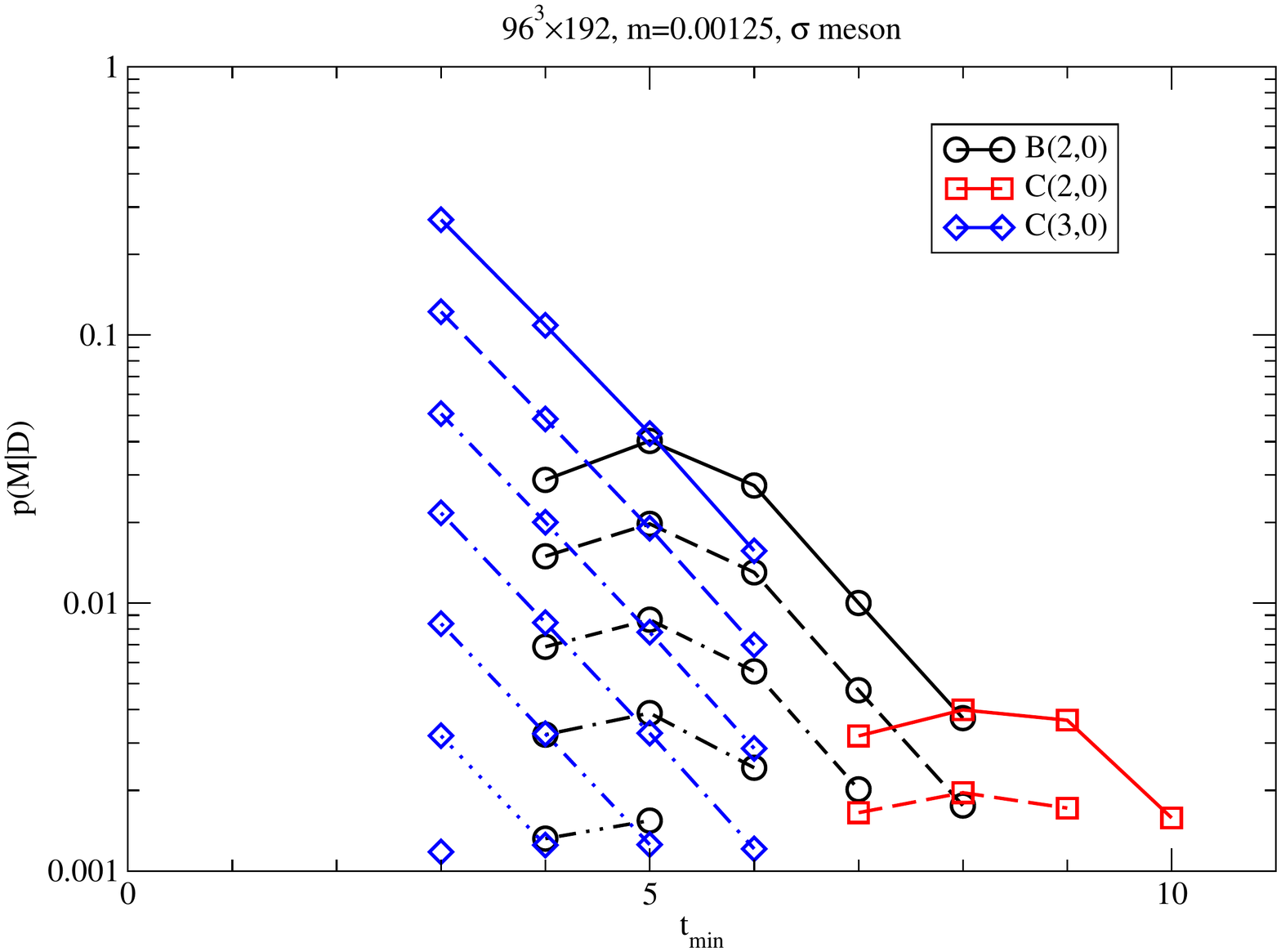}
\includegraphics[width=0.32\textwidth]{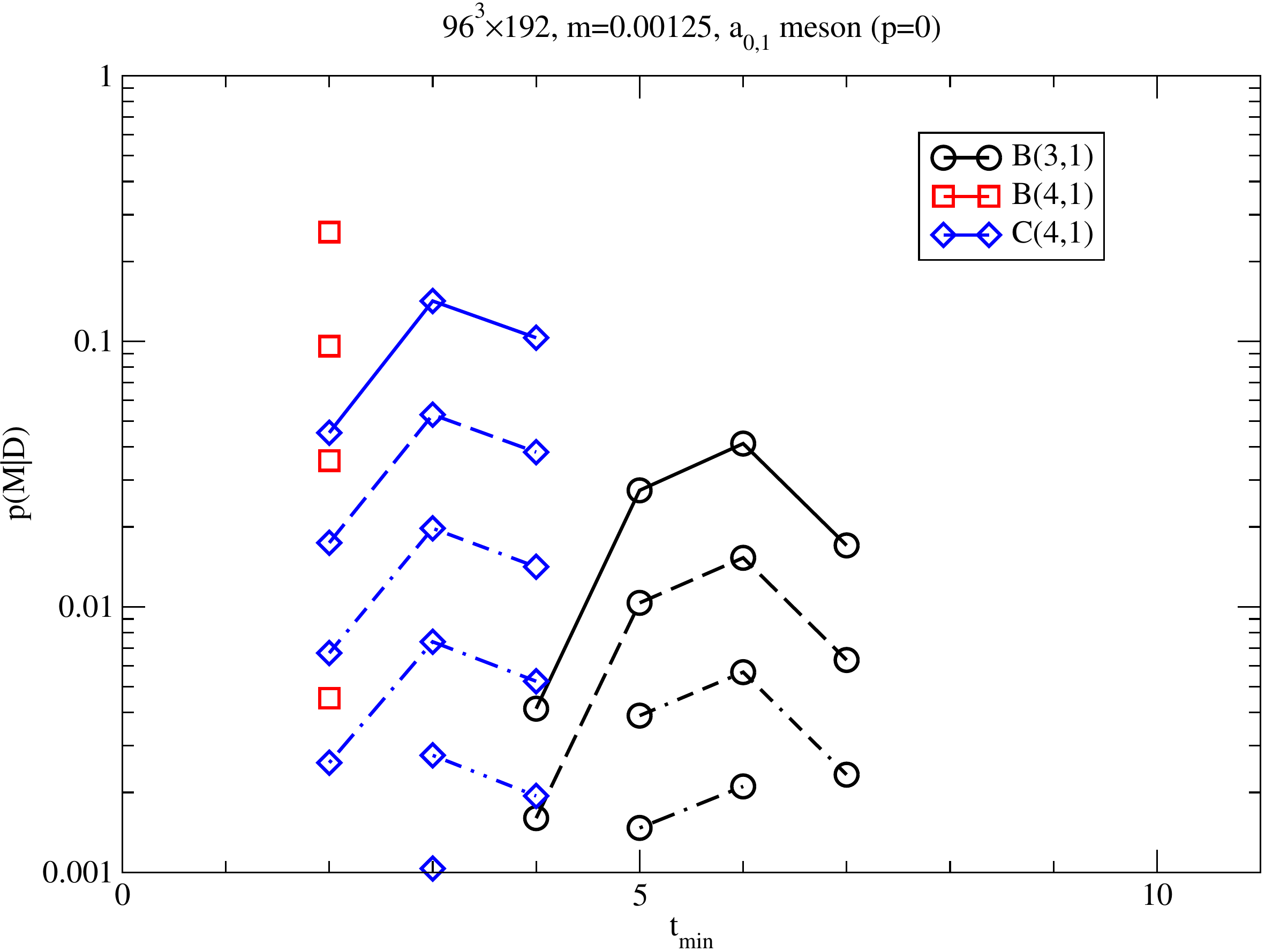}
\caption{\label{fig:p=0_MA} Relative model probabilities for the $\vec{p}=0$ $\pi_5$
$\sigma$ and $a_{0,1}$ mesons. The different models are labeled by a letter
A, B, C  and integers $(n_{\text{max}}, j_{\text{max}})$, the number of non-oscillating
and oscillating states, as described in Sec.~\ref{sub:model_function}.
The range of time values in each fit $[t_{\text{min}}, t_{\text{max}}]$ are shown
in the figures. The uppermost curves correspond to $t_{\text{max}}$ as the maximum value
in Tab.~\ref{tab:summ_t_range} and the lower curves correspond to decreasing $t_{\text{max}}$
by one.}
\end{figure}

As previously discussed, this analysis will use model averaging \cite{Jay:2020jkz}.
In Fig.~\ref{fig:p=0_MA}, we show how varying the fitting range
$t \in [t_{\text{min}}, t_{\text{max}}]$ affects the relative model probabilities
$p(M|D)$. We focus on the $\vec{p}=0$ mesons since those states are most susceptible
to the presence of $t$-invariant constant contribution to the correlation function.
This is true even in the case of the $a_{0,1}$ meson where the expected
constant contribution should vanish in the infinite statistics limit.
The $\pi_5$ meson is much less affected by any such constant as can be seen
by the preference for model A fits in the model averaging.

\subsection{Dispersive Analysis}

Once the model parameters and their errors have been computed for each correlation
function computed on a given volume, at a given fermion mass, and at a given
spatial momentum $\vec{p}$, the results from various momenta can be used
to constrain the values of the parameters in the rest frame using the dispersion
relations outlined in Eqs.~(\ref{eq:cont_disp}) through (\ref{eq:lat_disp})
for the rest mass $M_Q$ and Eqs.~(\ref{eq:F_pi_5_hat}) through
(\ref{eq:F_sigma_hat}) for the decay constants $\widehat{F}_Q$.
Parameter estimation is done using least-squared fitting with possible
finite lattice spacing corrections included in even powers of $\widehat{p}^2$
or $(a p)^2$, as appropriate.  Since the number of lattice correction terms needed
is unknown \textit{a priori} we use model averaging to average over the different
model choices.

\begin{figure}
\centering
\includegraphics[width=0.49\textwidth]{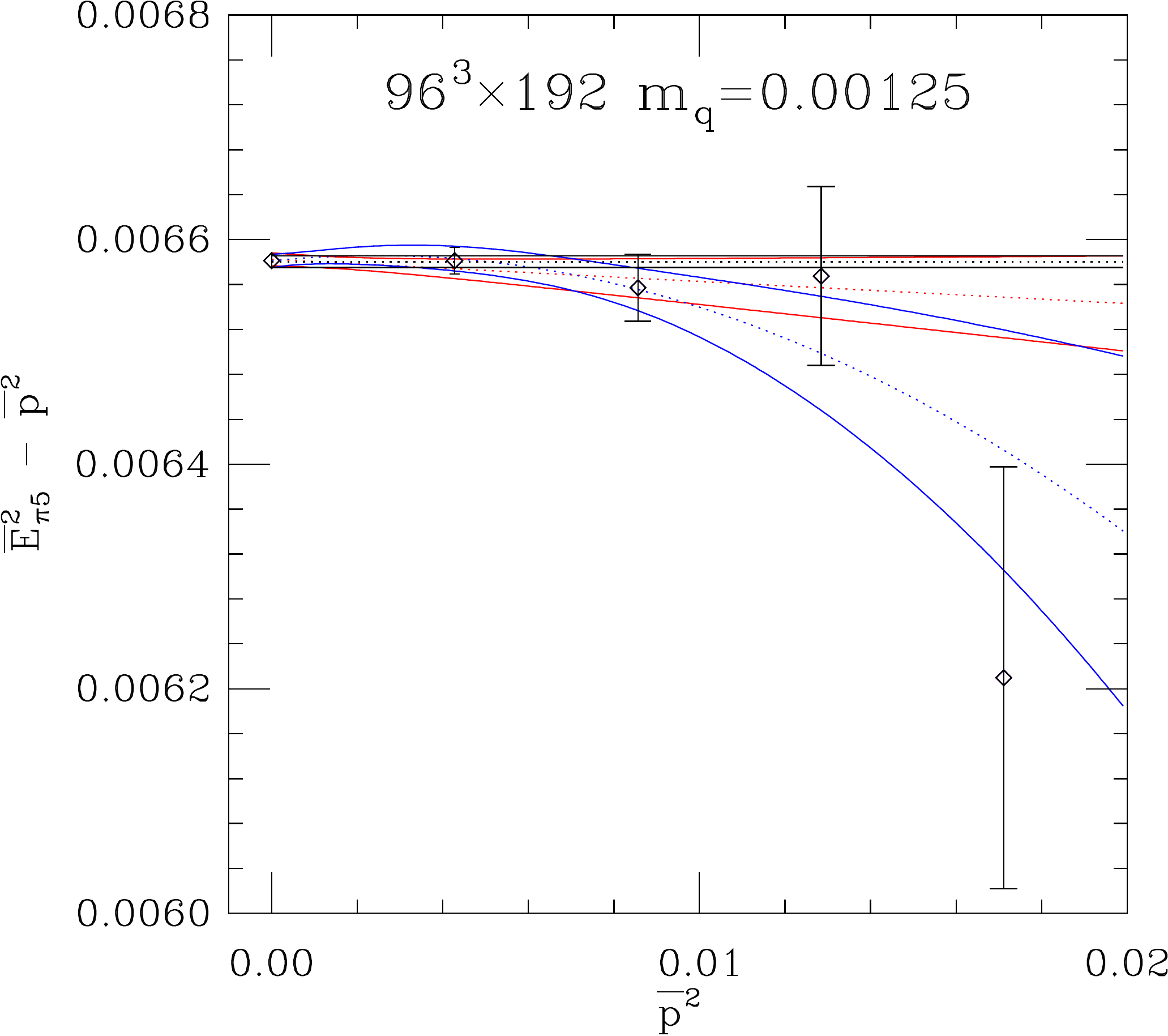}
\includegraphics[width=0.49\textwidth]{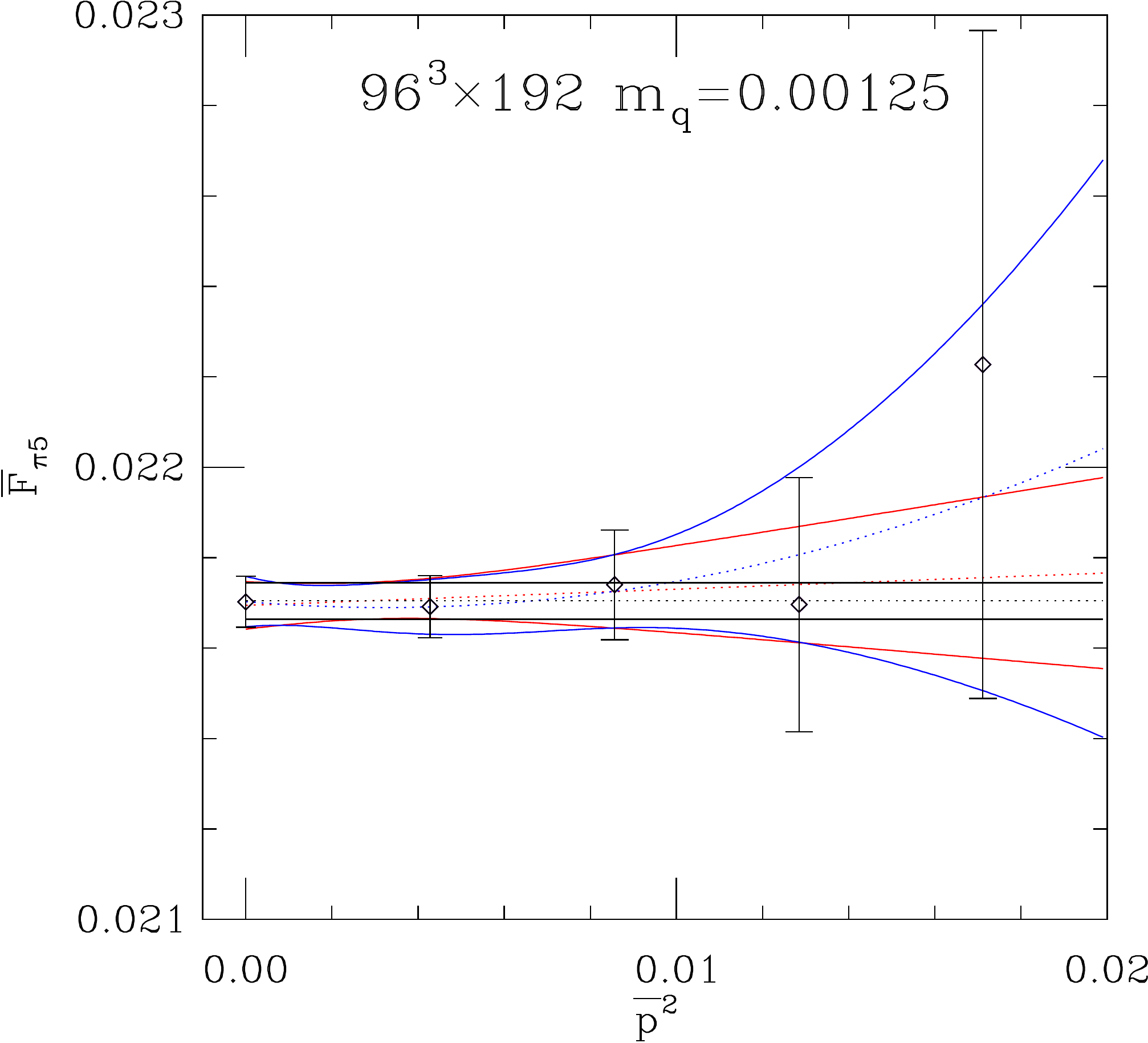}
\caption{Momentum dependence of the energy $\widehat{E}_{\pi_5}$ and
decay constant $\widehat{F}_{\pi_5}$. Fits to polynomials in $\widehat{p}^2$
up to quadratic order are shown.}
\label{fig:dispP}
\end{figure}

\begin{figure}
\centering
\includegraphics[width=0.49\textwidth]{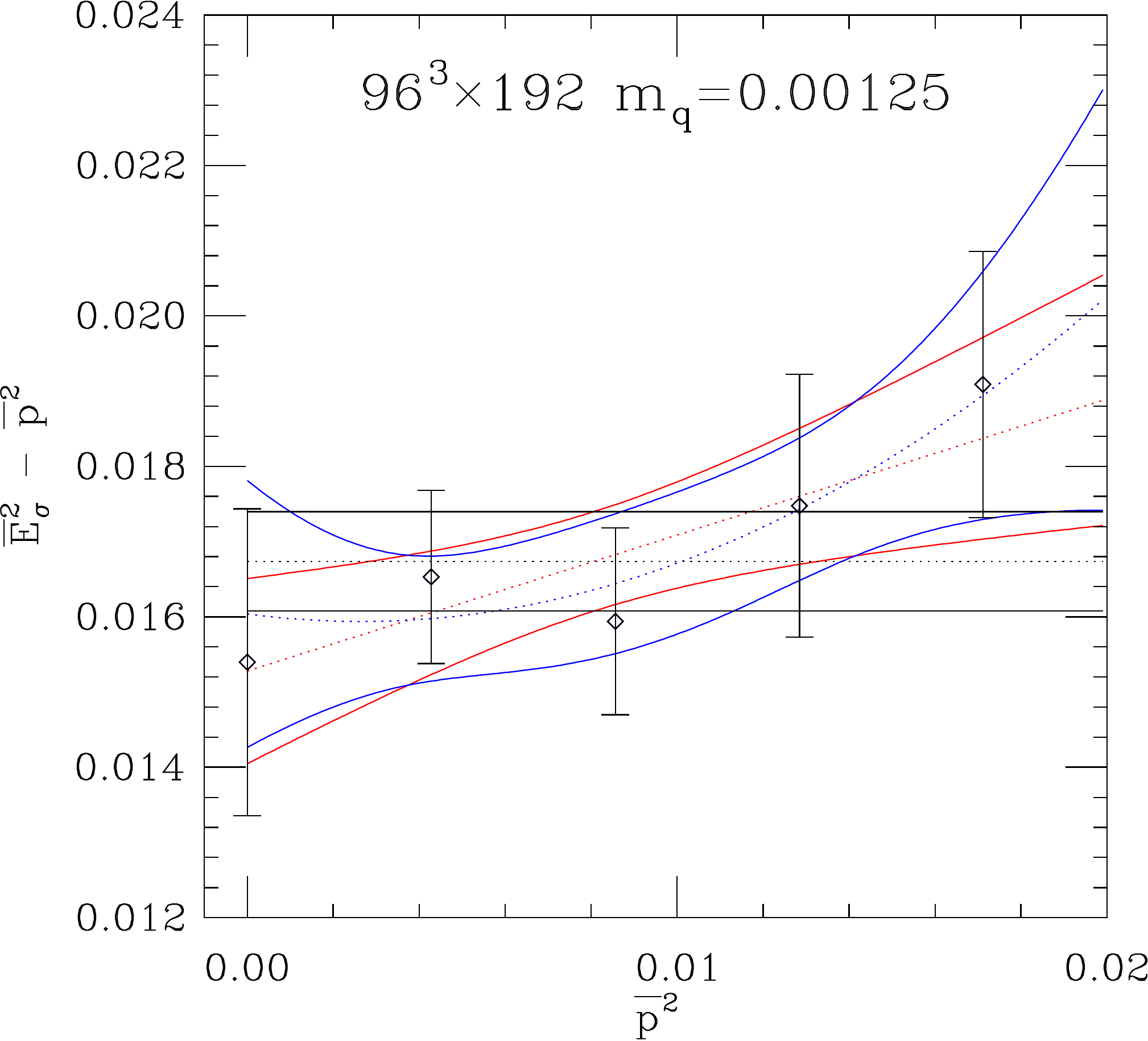}
\includegraphics[width=0.49\textwidth]{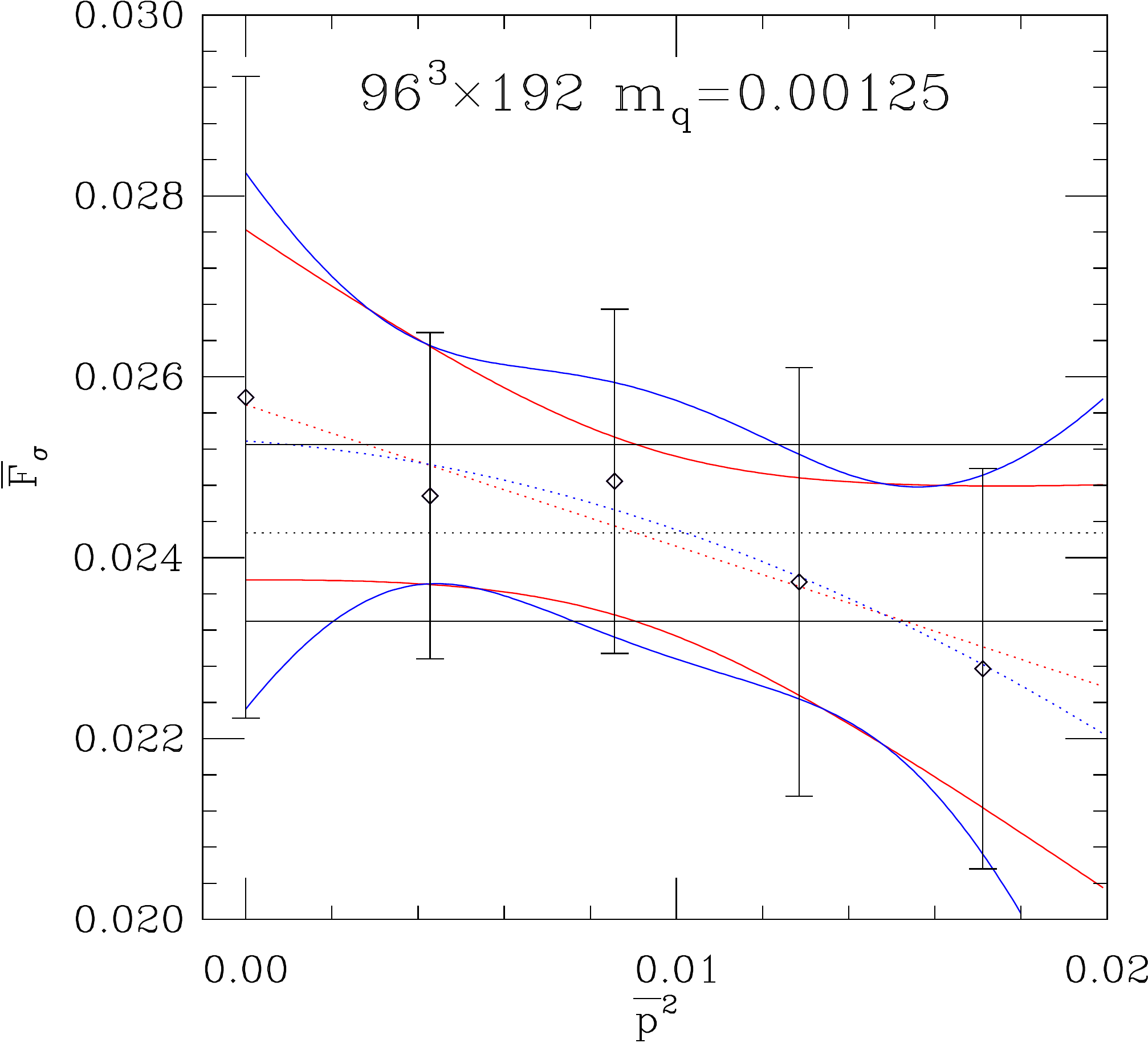}
\caption{Momentum dependence of the energy $\widehat{E}_{\sigma}$ and
decay constant $\widehat{F}_S$. Fits to polynomials in $\widehat{p}^2$
up to quadratic order are shown.}
\label{fig:dispS}
\end{figure}

\begin{figure}
\centering
\includegraphics[width=0.49\textwidth]{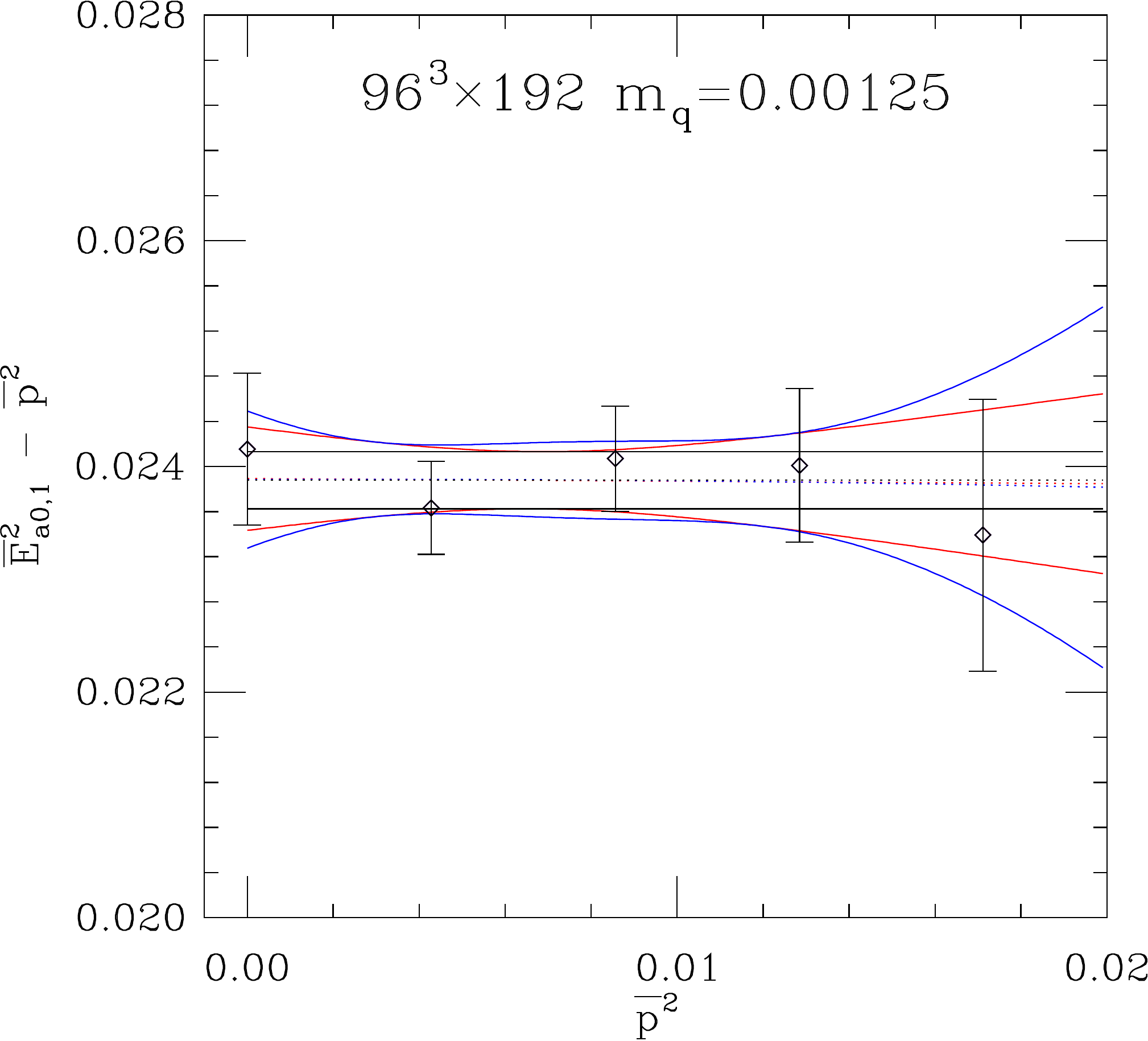}
\includegraphics[width=0.49\textwidth]{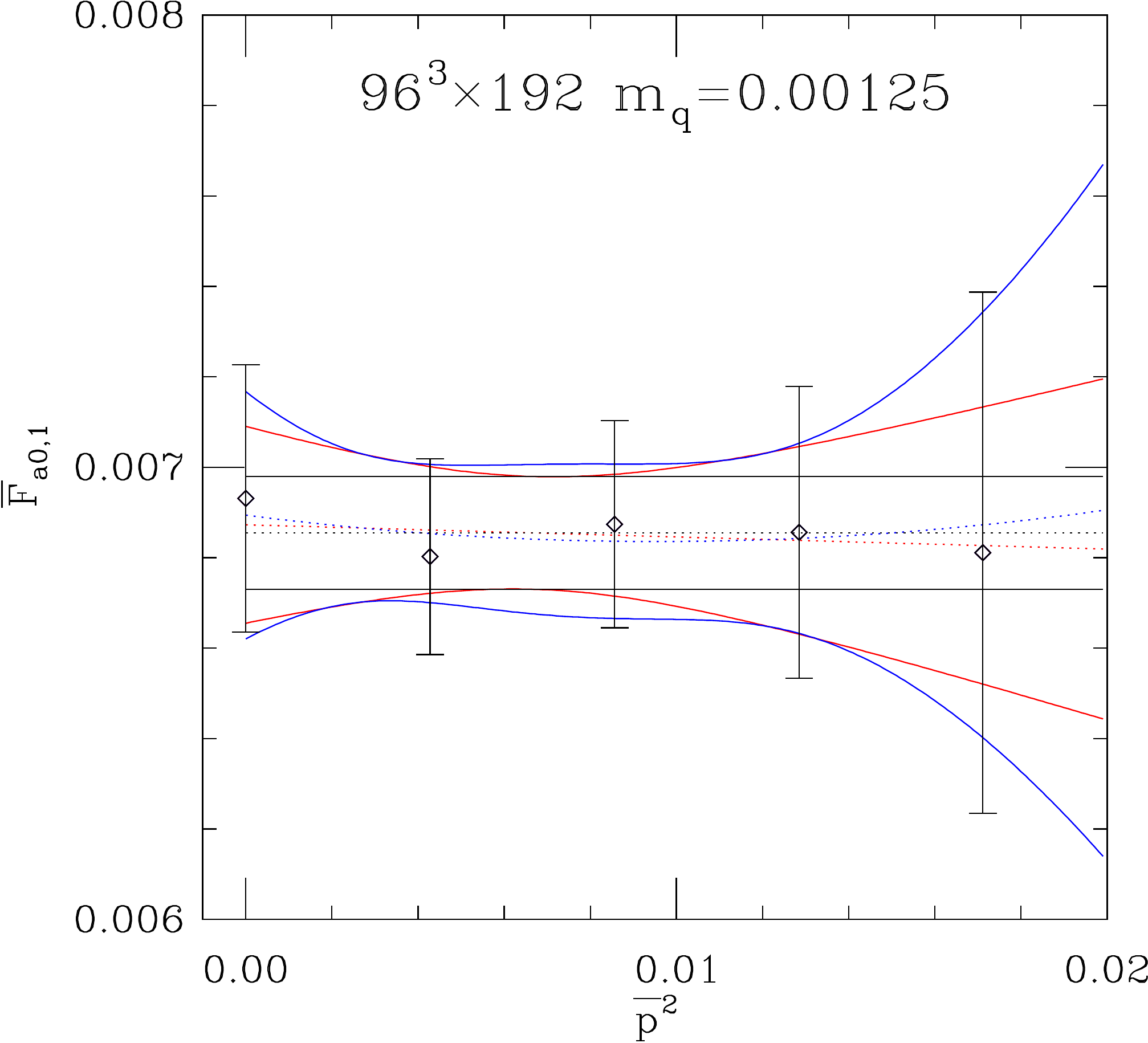}
\caption{Momentum dependence of the energy $\widehat{E}_{a_{0,1}}$ and
decay constant $\widehat{F}_{a_{0,1}}$.}
\label{fig:dispC}
\end{figure}

This procedure is probably of marginal benefit for the $\pi_{5}$ rest mass
and decay constant since those quantities are already very accurately
determined directly in the $\vec{p}=0$ frame and the other momentum frames
do not add significant additional information, as shown in Fig.~\ref{fig:dispP}.
However, these fits also show how the momentum-dependence is consistent
with the expected dispersion relations up to small lattice artifacts.

For the isosinglet scalar $\sigma$ in Fig.~\ref{fig:dispS} and isotriplet scalar
$a_{0,1}$ in Fig.~\ref{fig:dispC}, we see similar consistency with the expected
dispersion relations.  Now, the information from the non-zero momentum
frames provides additional significant constraints on the rest mass and decay
constant resulting in overall smaller uncertainties than if only the $\vec{p}=0$
results alone were used.  This is particularly important for the $\sigma$ channel
where the correlation function in the $\vec{p}=0$ frame has a difficult
to subtract constant which is not present in non-zero momentum frames.

\section{Infinite volume extrapolation}

We repeat the steps described in detail for one ensemble in
Sec.~\ref{sec:detailed_example} for all ensembles in this study.
We would like to compare the results of our calculations with various models
but those models usually only apply to the system in an infinite volume.
We will extrapolate our data to the infinite volume limit using the model
described in Sec.~\ref{sub:finite_volume}.  At each volume and fermion mass
we compute the quantity
\begin{equation}
\xi(m_q, L) \equiv \frac{M_{\pi_5}^2}{(4 \pi \widehat{F}_{\pi_5})^2}
\sum_{n=1}^8 \frac{4 \ \kappa(n)}{\sqrt{n} \ M_{\pi_5} L}
K_1(\sqrt{n} \ M_{\pi_5} L)
\end{equation}
where the $m_q$ dependence is implicit in the relevant infinite volume
quantities $M_{\pi_5}$, $\widehat{F}_{\pi_5}$.  With this computed quantity,
the analysis becomes a simple linear fit.

If we focus just on $M_{\pi_5}$ and $\widehat{F}_{\pi_5}$,
we know in chiral perturbation theory, the quantities $\alpha_{\pi_5}$
and $\beta_{\pi_5}$ defined in Eqs.~(\ref{eq:M_X_FV}) and (\ref{eq:F_X_FV})
appear at a specific order in the chiral expansion and have no implicit fermion mass
dependence.  We use the same finite volume model for other masses and decay
constants and we will similarly assume the parameters $\alpha_Q$, $\beta_Q$
are mass-independent as a model choice.  This means that $\alpha_Q$, $\beta_Q$
are determined by a simultaneous fit to the data at all fermion masses and volumes.


The choice of the expansion parameter $\xi(m_q, L)$ being defined
in terms of infinite-volume quantities might pose a chicken-and-egg problem
when attempting to extrapolate $\pi_5$ data since the infinite volume values
are not known \textit{a priori}.  In this case, we start by using the values
on the largest volume and then iterate a few times
and the result converges quickly.

\begin{figure}
\centering
\includegraphics[width=0.49\textwidth]{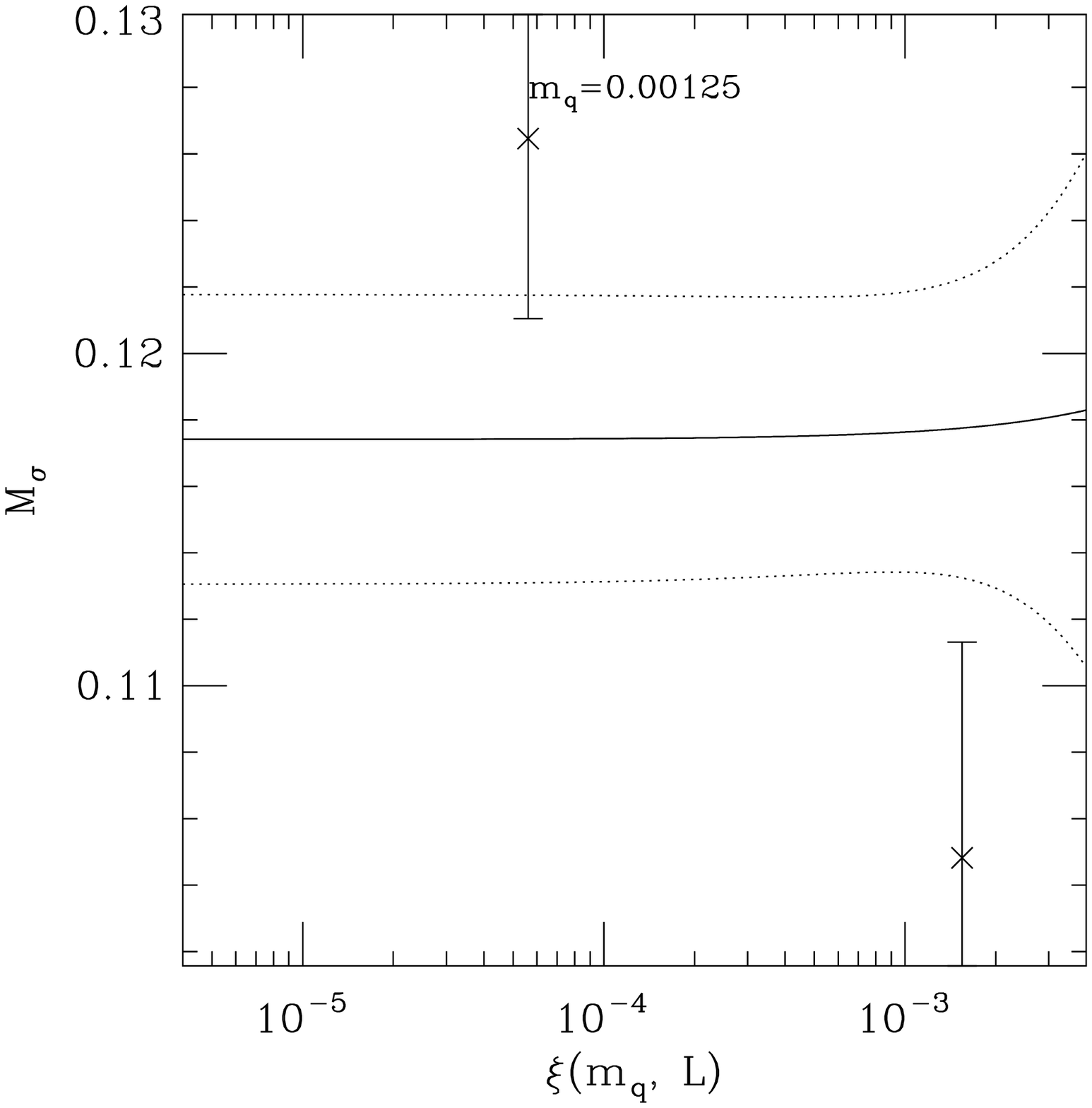}
\includegraphics[width=0.49\textwidth]{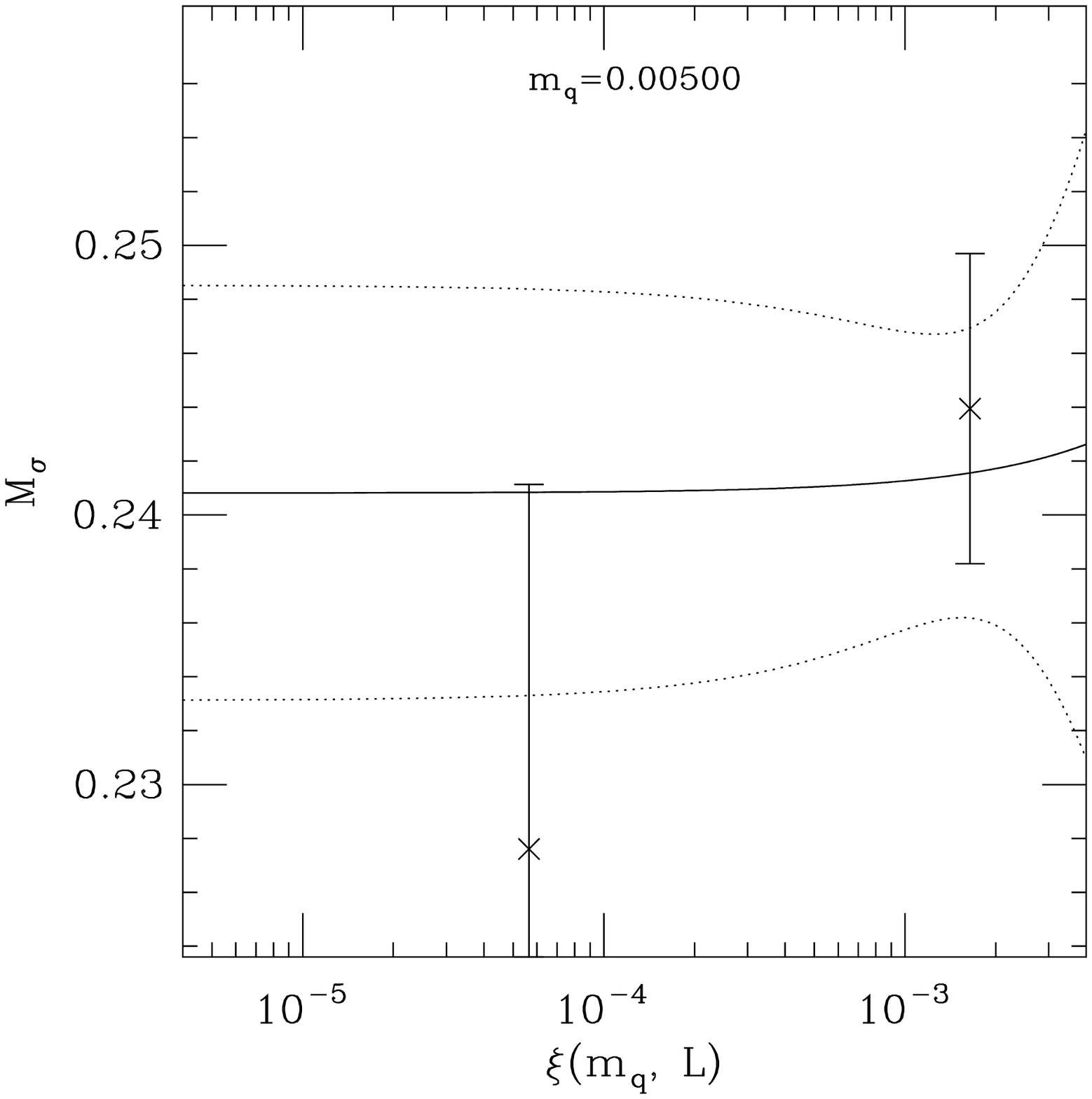} \\
\includegraphics[width=0.49\textwidth]{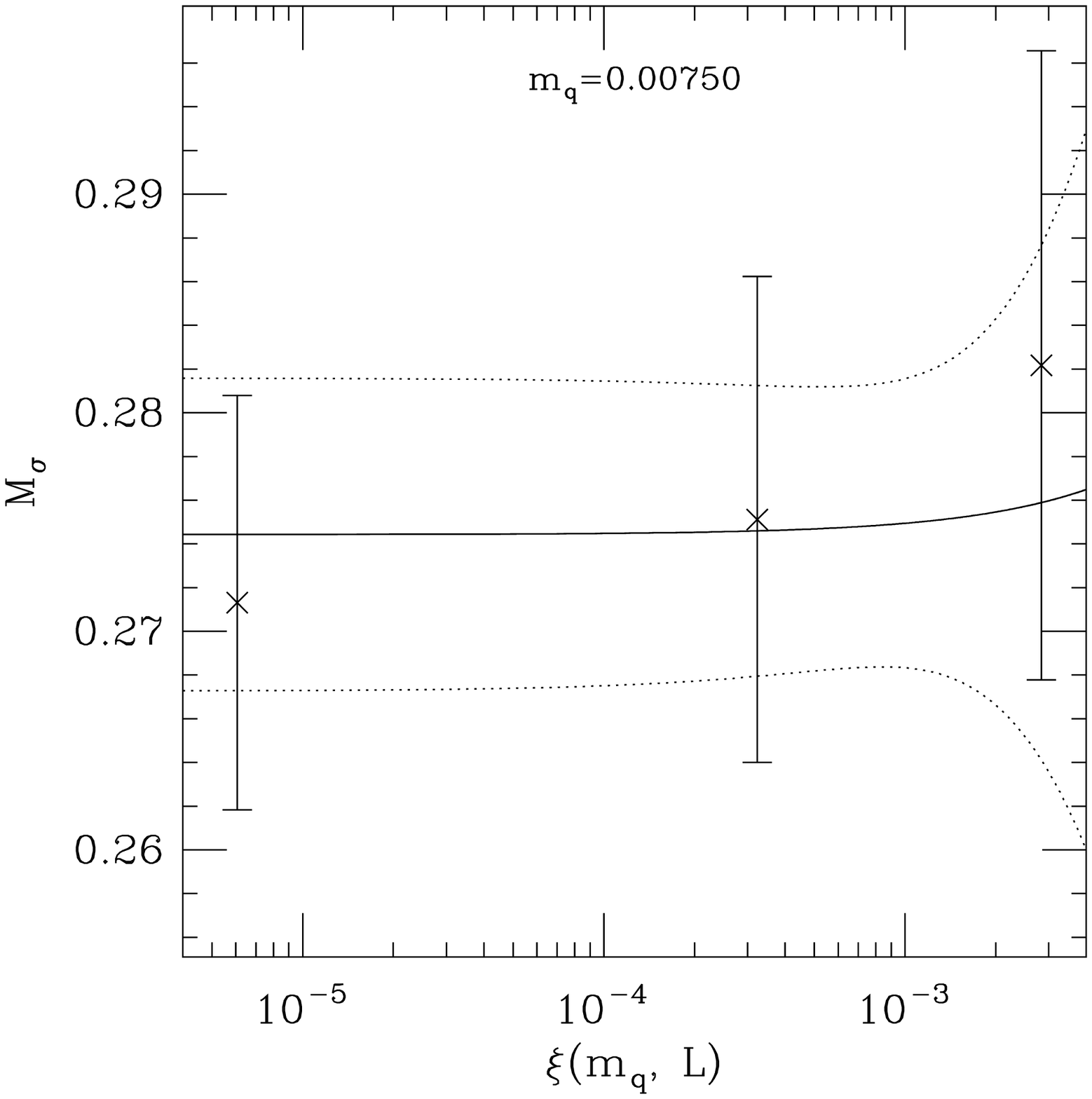}
\includegraphics[width=0.49\textwidth]{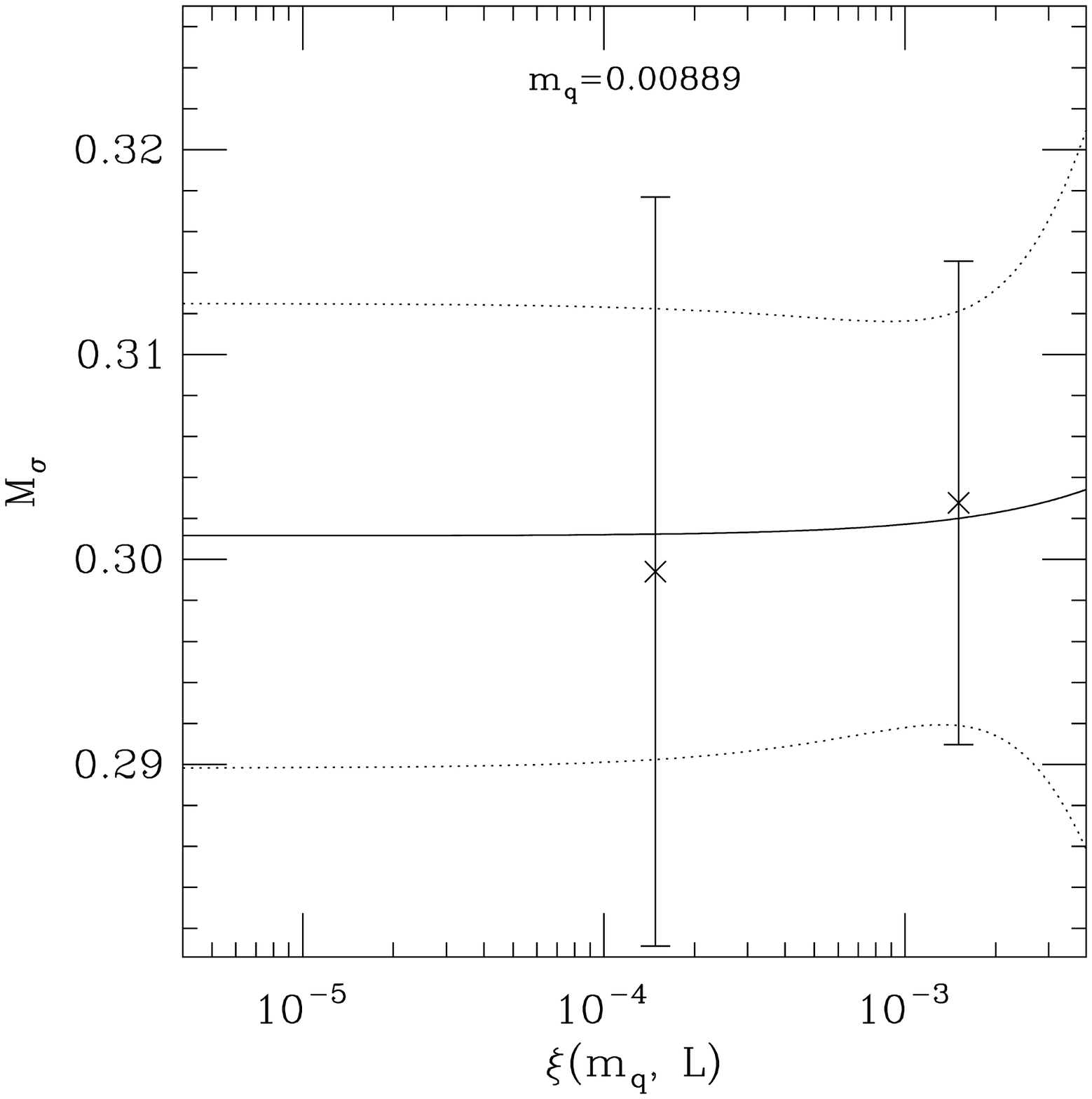}
\caption{Infinite volume extrapolation of the rest masses $M_\sigma(m_q)$.
$\alpha_\sigma = 1.9 \pm 16.7$ and $\chi^2/\text{dof}=2.05$ with 4 dof.}
\label{fig:MS_vs_xi}
\end{figure}

\begin{figure}
\centering
\includegraphics[width=0.49\textwidth]{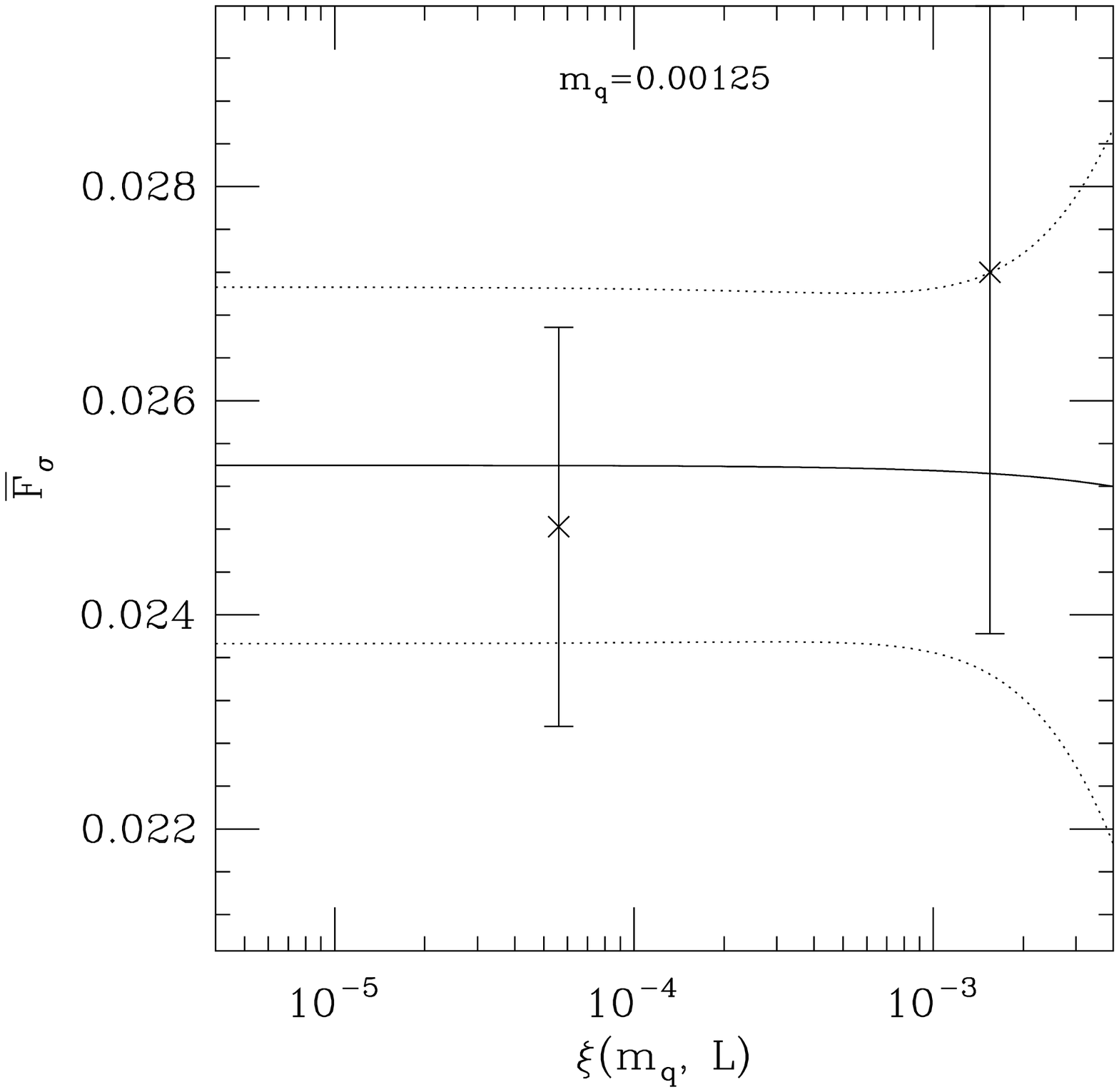}
\includegraphics[width=0.49\textwidth]{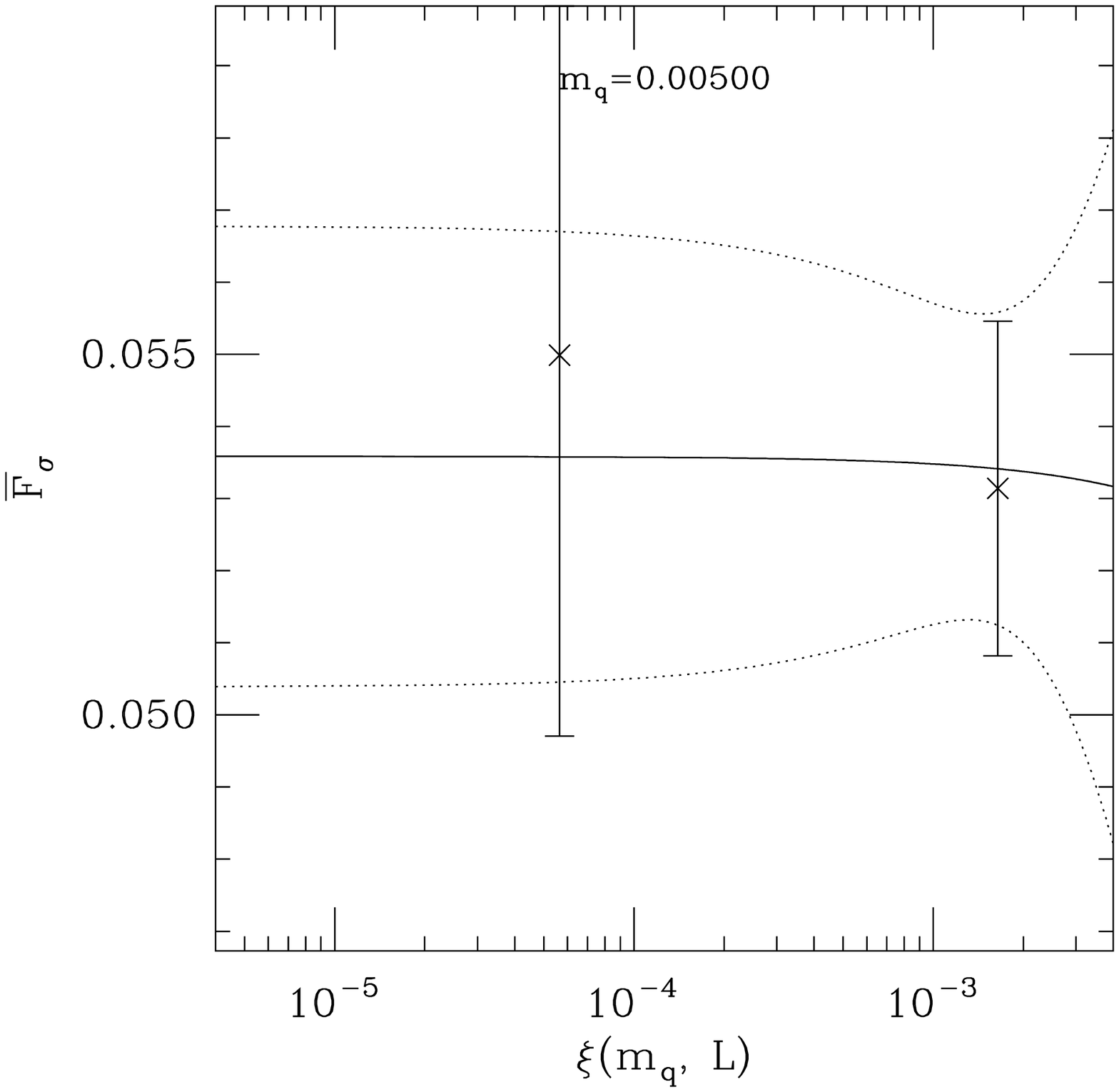} \\
\includegraphics[width=0.49\textwidth]{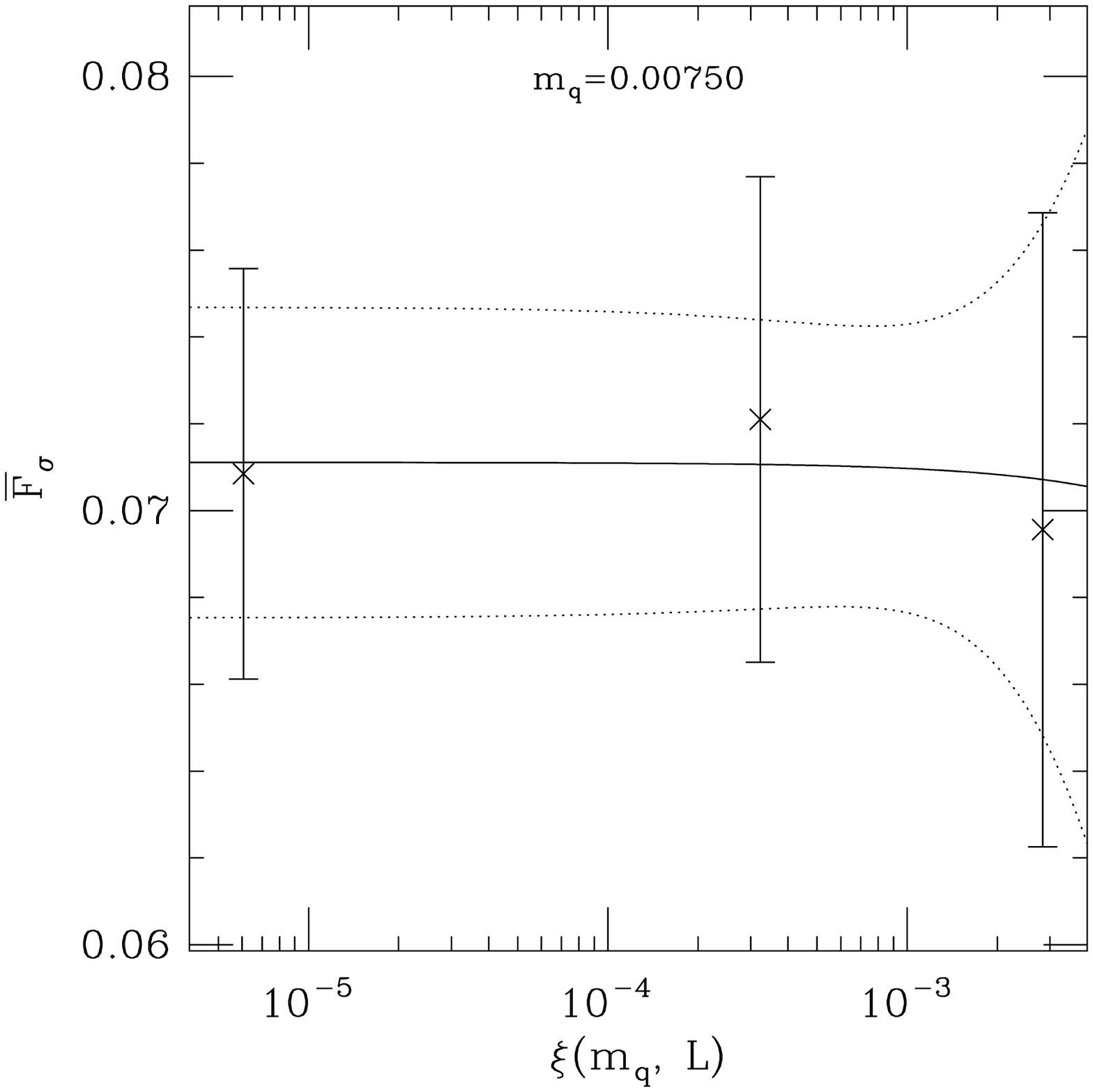}
\includegraphics[width=0.49\textwidth]{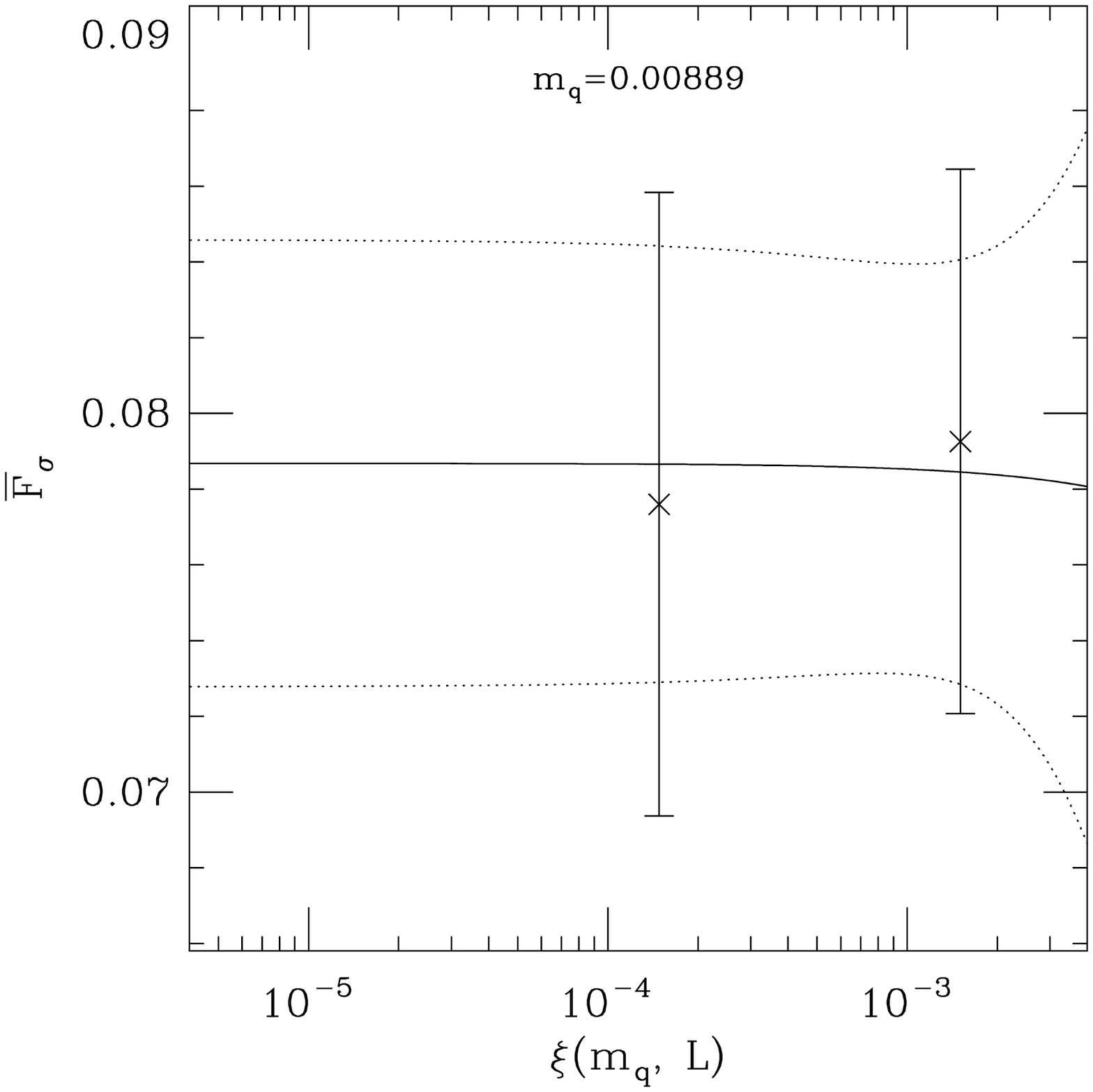}
\caption{Infinite volume extrapolation of the decay constants
$\widehat{F}_S(m_q)$. $\beta_\sigma = -1.9 \pm 32.0$ and $\chi^2/\text{dof}
= 0.15$ with 4 dof.}
\label{fig:FS_vs_xi}
\end{figure}

An earlier version of the finite volume extrapolation for $M_{\pi_5}$
and $F_{\pi_5}$ were published previously \cite{LatticeStrongDynamics:2021gmp}
where it was observed to be a relatively minor correction on our volumes.
Our current results are consistent with them, so we focus here on $\sigma$
channel.  The fit of $M_\sigma(m_q, L)$ is shown in Fig.~\ref{fig:MS_vs_xi}
and the fit of $\widehat{F}_S(m_q, L)$ is shown in Fig.~\ref{fig:FS_vs_xi}.
$\alpha_Q, \beta_Q$ for various channels studied in this work are summarized in 
Table~\ref{tab:alpha_beta}. Both from the figures and from the uncertainties
on $\alpha_\sigma$ and $\beta_\sigma$ in the table, it is clear that the uncertainties
in our $\sigma$ meson observables are still too large to reliably extract the sign
and magnitude of these finite volume corrections.  We hope to return to this issue
in a future publication.

Studying the other parameters in Table~\ref{tab:alpha_beta}
reveals relationships between
parameters which are generated by the strong dynamics and
which are qualitatively similar to QCD.  First, $\text{sign}(\alpha_Q)
= - \text{sign}(\beta_Q)$ is a well-known feature in QCD.  Second, the fact
that $\text{sign}(\alpha_{\pi_5}) = - \text{sign}(\alpha_{a_{0,1}})$ is 
also observed in earlier studies \cite{Fleming:2008gy} and was previously
misinterpreted as an indication of ``parity doubling'' in near-conformal gauge theories
because finite volume effects would push the masses and decay constants of parity
partners $\pi_5$ and $a_{0,1}$ towards degeneracy. We also note that without a proper
infinite-volume extrapolation, if the mass of the $a_{0,1}$ meson were observed
to be stable but just below decay threshold, one could wonder whether the state might
become unstable in a larger volume.  In our calculations,
the $a_{0,1}$ meson remains stable even after infinite volume extrapolation
as can be seen in Tab.~\ref{tab:data_stat_only}.

\begin{table}[ht]
\centering
\addtolength{\tabcolsep}{3 pt}   
\begin{tabular}{c||c|c||c|c}
  & $\alpha_Q$ & $\chi^2/\text{dof}$ & $\beta_Q$ & $\chi^2/\text{dof}$ \\
\hline
$\pi_5$   & 6.53(29)   & 3.12 & -9.3(1.3) & 0.65 \\
$\sigma$  & 2(17)      & 2.05 & -2(32)    & 0.15 \\
$a_{0,1}$ & -27.4(4.6) & 2.88 & 11(11)    & 0.29 \\
\end{tabular}
\caption{Summary of finite volume corrections $\alpha_Q, \beta_Q$. All fits
have 4 dof.  Multiply these parameters by $3/(2 \sqrt{2 \pi^3}) \approx 0.19$
to compare with \cite{LatticeStrongDynamics:2021gmp}.}
\label{tab:alpha_beta}
\end{table}

\begin{table}[ht]
  \centering
  \addtolength{\tabcolsep}{3 pt}   
  \begin{tabular}{c||l|l|l|l|l}
    $m_q$ & 0.00125 & 0.00222 & 0.00500 & 0.00750 & 0.00889 \\
    \hline
    $M_{\pi_5}$ &
    0.081082(32) & 0.10870(12) & 0.165691(73) & 0.205711(33) & 0.22534(13) \\
    $\widehat{F}_{\pi_5}$ &
    0.021677(40) & 0.02794(12) & 0.03982(10) & 0.048314(66) & 0.05262(15) \\
    $M_\sigma$ &
    0.1174(44) & 0.1545(79) & 0.2408(77) & 0.2744(71) & 0.301(11) \\
    $\widehat{F}_S$ &
    0.0254(17) & 0.0361(37) & 0.0536(32) & 0.0711(36) & 0.0787(59) \\
    $M_{a_{0,1}}$ &
    0.1536(10) & 0.2070(53) & 0.3119(28) & 0.3773(18) & 0.4193(32) \\
    $\widehat{F}_{a_{0,1}}$ &
    0.00691(14) & 0.00944(43) & 0.1480(27) & 0.01829(24) & 0.02047(42) \\
    $M_{\rho_i}$ \cite{LatticeStrongDynamics:2018hun} &
    0.1709(65) & 0.2197(37) & 0.3024(63) & 0.36962(77) & 0.4093(21)
  \end{tabular}
  \caption{\label{tab:data_stat_only}Final summary of infinite volume ground
  state rest masses and decay constants in lattice units. Only statistical
  uncertainties are shown.  Data for $M_{\rho_i}$ copied from
  \cite{LatticeStrongDynamics:2018hun} for convenience.
  See Table~\ref{tab:data_stat+sys} for results with systematic
  uncertainties included. Supplementary results for all fit parameters
  are available \cite{LatticeStrongDynamics:2023zenodo}.}
\end{table}


\section{\label{sec:systematics}Systematic Error Analysis}

In our previous $I=2, \pi\pi$ scattering paper
\cite{LatticeStrongDynamics:2021gmp}, we made a crude estimate of the relative
systematic errors affecting our statistical determinations of the $\pi_5$
meson mass $M_{\pi_5}(m_q)$ and decay constant $F_{\pi_5}(m_q)$.
Our current
statistical-only estimate of uncertainties for quantities like $M_{\pi_5}(m_q)$
and $\widehat{F}_{\pi_5}(m_q)$ as shown in Tab.~\ref{tab:data_stat_only} are likely
underestimates due to various factors: a small
number of independent samples; various modeling choices regarding dispersion relations
and  finite volume effects; data quality cuts and model probability cuts in the model
averaging procedure; plus the interplay between the amount of independent data
and choices made in the rest of the analysis through the reliability of the shrinkage
estimator of covariance.

We would like to estimate how large these effects might be in terms of a single relative
systematic error parameter $\rho$ across all the ensembles.  We will estimate $\rho$
using a number of different observables and then combine those estimates to get
an average value for $\rho$.  For example,
if $\sigma_M$ is the statistical-only estimate of the uncertainty of a given mass
$M$, we would like to estimate a relative systematic uncertainty $\rho_M$
such that the total uncertainty is
\begin{equation}
\label{eq:total_uncertainty}
M(m_q) \pm \sqrt{\sigma_{M(m_q)}^2 + M(m_q)^2 \rho_M^2}
\end{equation}
We assume that the systematic effect is similar across all the different
ensembles labeled by different fermion masses $m_q$ so that the parameter $\rho_M$
doesn't depend on $m_q$.

To estimate $\rho_M$, we don't want to assume any
explicit functional dependence for $M(m_q)$, in particular, that we would expect
to be valid for small $m_q$ including as $m_q \to 0$.
Instead, we imagine that whatever
the correct function, it is relatively smooth and slowly-varying and can be
approximated by a Taylor series expansion around the midpoint $m_0 = 0.00507$
of our range of $m_q$ and $| m_q - m_0 | \le \Delta_m = 0.00382$. We can fit
the data to a polynomial
\begin{equation}
\label{eq:power_series}
M(m_q) \approx \sum_{n=0}^{n_\text{max}} a_n \left( m_q - m_0 \right)^n
\end{equation}
Given that we have only five different $m_q$ values, we will compare the $\chi^2$
and AIC values for $n_\text{max}$=2 and $n_\text{max}$=3 and use those comparisons
to estimate $\rho_M$.
We will also use
the ratio test to check for convergence of the series on $m_0 \pm \Delta_m$
\begin{equation}
\label{eq:ratio_test}
\frac{|a_{n+1} \Delta^{n+1}|}{|a_{n} \Delta^n |} < 1, \quad \forall n
\end{equation}
Actually the ratio test only requires the ratio $< 1$ as $n \to \infty$
for convergence, but we will assume convergence if its true term-by-term up
to the largest $n$ we can fit. For this analysis, we will use the data
in Tab.~\ref{tab:data_stat_only}.

\subsection{Fits using statistical-only data}

In this section, in Tab.~\ref{tab:basic_fits} we show fits of Eq.~(\ref{eq:power_series})
to the statistical-only data from Tab.~\ref{tab:data_stat_only} for $n_\text{max}$=2, 3.
We then test for convergence by computing the ratios in Eq.~(\ref{eq:ratio_test})
and collect the results in Tab.~\ref{tab:basic_ratio_test}.

\begin{table}[h]
\centering
\addtolength{\tabcolsep}{3 pt}   
\begin{tabular}{l|l||l|l|l|l|l|l}
obs. & $n_\text{max}$ & $\chi^2$ & $\log p(M|D)$ & $a_0$ & $a_1$ & $a_2$ & $a_3$ \\
\hline\hline
$M_{\pi_5}$
  & 2 & 1052.0 & -528.0 & 0.166768(60) & 18.563(11) & -1000.9(6.4) & \\
  & 3 &  159.5 &  -83.8 & 0.167345(63) & 17.472(38) &  -950.4(6.6) & 100500(3400) \\
\hline
$\widehat{F}_{\pi_5}$
  & 2 & 76.0 & -41.0 & 0.040144(82) &  3.967(15) & -222.3(8.4) & \\
  & 3 & 14.3 & -11.2 & 0.040275(84) &  3.581(51) & -207.4(8.6) & 33600(4200) \\
\hline
$M_{\sigma}$
  & 2 & 1.6 & \textbf{\color{red}-3.8} 
                   & 0.2382(64) & 22.9(1.3) & -2250(640) & \\
  & 3 & 0.8 & \textbf{\color{red}-4.4} 
                   & 0.2384(64) & 19.5(4.1) & -2122(657) & 29(33) $\times 10^4$ \\
\hline
$\widehat{F}_S$
  & 2 & 0.6 & \textbf{\color{red}-3.3} 
                   & 0.0551(28) & 6.93(62) & -200(290) & \\
  & 3 & 0.4 & \textbf{\color{red}-4.2}
                   & 0.0550(28) & 6.1(2.0) & -160(300) & 6(16) $\times 10^4$ \\
\hline
$M_{a_{0,1}}$
  & 2 & 14.7 & -10.4 & 0.3097(22) & 33.82(36) & -1820(210) & \\
  & 3 &  0.0 &  -4.0 & 0.3136(25) & 28.0(1.5) & -1870(210) & 46(12) $\times 10^4$ \\
\hline
$\widehat{F}_{a_{0,1}}$
  & 2 & 1.9 & \textbf{\color{red}-3.9}
                   & 0.01483(23) & 1.726(46) & -89(23) & \\
  & 3 & 0.0 & \textbf{\color{red}-4.0}
                   & 0.01489(23) & 1.50(17)  & -82(23) & 19(14) $\times 10^3$ \\
\hline
\end{tabular}
\caption{\label{tab:basic_fits} Basic fits using statistical-only data
from Tab.~\ref{tab:data_stat_only} to model function in Eq.~(\ref{eq:power_series}).
Bolded entries indicate observables where
model probabilities are higher for $n_\text{max}=2$ than $n_\text{max}=3$.}
\end{table}

\begin{table}[h]
\centering
\addtolength{\tabcolsep}{3 pt}   
\begin{tabular}{l|l||l|l|l}
obs. & $n_\text{max}$ & \multicolumn{1}{c|}{$\Delta_m \left|\frac{a_1}{a_0}\right|$}
  & \multicolumn{1}{c|}{$\Delta_m \left|\frac{a_2}{a_1}\right|$}
  & \multicolumn{1}{c }{$\Delta_m \left|\frac{a_3}{a_2}\right|$} \\
\hline\hline
$M_{\pi_5}$
  & 2 & 0.42521(36) & 0.2060(14) & \\
  & 3 & 0.39884(95) & 0.2078(15) & 0.404(15) \\
\hline
$\widehat{F}_{\pi_5}$
  & 2 & 0.3775(19) & 0.2141(86) & \\
  & 3 & 0.3397(56) & 0.2212(97) & 0.618(88) \\
\hline
$M_{\sigma}$
  & 2 & 0.368(24) & 0.37(12) & \\
  & 3 & 0.312(68) & 0.42(15) & \textbf{\color{red}0.52(66)} \\
\hline
$\widehat{F}_S$
  & 2 & 0.481(54) & 0.11(16) & \\
  & 3 & 0.43(14) & 0.10(19) & \textbf{\color{red}1.6(5.9)} \\
\hline
$M_{a_{0,1}}$
  & 2 & 0.4171(64) & 0.206(26) & \\
  & 3 & 0.342(20) & 0.254(35) & \textbf{\color{red}0.94(26)} \\
\hline
$\widehat{F}_{a_{0,1}}$
  & 2 & 0.445(15) & 0.197(54) & \\
  & 3 & 0.358(46) & 0.211(64) & \textbf{\color{red}0.86(72)} \\
\hline
\end{tabular}
\caption{\label{tab:basic_ratio_test} Ratios for convergence testing of fits
in Tab.~\ref{tab:basic_fits}. The fit parameter
covariance matrix (not shown) was used to compute these uncertainties.
Bolded entries indicate observables where the convergence test may fail
due to large values or uncertainties.}
\end{table}

\begin{table}[h]
\centering
\addtolength{\tabcolsep}{3 pt}   
\begin{tabular}{l|l||l|l}
obs. & $n_\text{max}$ & $f^\prime(m_q)=0$ & $f^{\prime\prime}(m_q)=0$ \\
\hline\hline
$M_{\pi_5}$
  & 2 & 0.013 & \\
  & 3 & $0.0070 \pm 0.0069 i$ & 0.0070 \\
\hline
$\widehat{F}_{\pi_5}$
  & 2 & 0.013 & \\
  & 3 & $0.0059 \pm 0.0056 i$ & 0.0059 \\
\hline
$M_{\sigma}$
  & 2 & 0.0089 & \\
  & 3 & $0.0062 \pm 0.0040 i$ & 0.0062 \\
\hline
$\widehat{F}_S$
  & 2 & 0.021 & \\
  & 3 & $0.0046 \pm 0.0054 i$ & 0.0046 \\
\hline
$M_{a_{0,1}}$
  & 2 & 0.0131 & \\
  & 3 & $0.0052 \pm 0.0043 i$ & 0.0052 \\
\hline
$\widehat{F}_{a_{0,1}}$
  & 2 & 0.0135 & \\
  & 3 & $0.0053 \pm 0.0050 i$ & 0.0053 \\
\hline
\end{tabular}
\caption{\label{tab:derivatives}Zeroes of the derivatives of fits
in Tab.~\ref{tab:basic_fits}.}
\end{table}

If we first look at the model probabilities, we see when the fit is highly-constrained,
indicated by large $\chi^2$ values, then the fit with $n_\text{max}=3$ is preferred relative
to $n_\text{max}=2$.  This is the expected behavior since adding extra fit parameters
in a highly-constrained fit usually reduces the $\chi^2$ by a sufficient amount
to increase the model probability.  However, if the fit is poorly-constrained,
indicated by a small $\chi^2$, adding extra parameters may not increase the model
probability.  Observables where this occurs are highlighted in
Table~\ref{tab:basic_fits}
and those observables are probably too noisy to help constrain the systematic error
parameter $\rho$.

Looking at the convergence test in Table~\ref{tab:basic_ratio_test}, again we highlight
examples where data were too noisy to pass the test with confidence.  Again, we will
not use those observables to help constrain $\rho$.  Note also the strong overlap
in the lists of rejected observables from both tables.  Finally, we don't expect that
the functions will have extremal points in the region where
it approximates the data.  The zeroes of the derivatives are shown
in Tab.~\ref{tab:derivatives}.

\subsection{Estimating Relative Systematic Error}

To estimate the relative systematic error parameter $\rho$,
from Eq.~(\ref{eq:total_uncertainty}) as $\rho$
increases the error bars will increase and the corresponding $\chi^2$
will decrease. What value of $\chi^2$ should we choose to determine $\rho$?
\textit{A priori}, two interesting values come to mind: \textbf{(I)} the mean
value of the chi-squared distribution for $k$ degrees of freedom,
\textit{i.e.}\ $\chi_k^2(\rho^{(I)}) = k$; \textbf{(II)} the value
of $\chi^2_k$ such that one expects 68\% of the time a random sample
of the chi-squared distribution should be less than or equal to that value,
\textit{i.e.} $\chi_1^2(\rho) = 1$, $\chi_2^2(\rho^{(II)}) \approx 2.3$.

\textit{A posteriori}, we noticed that from a model averaging perspective,
the 4-parameter cubic polynomial fit has the higher model probability
at $\rho$=0 in cases where the statistical error is small compared to the
expected systematic error.  In the $\rho \to \infty$ limit, $\chi^2 \to 0$
and the most likely model is the one with the smallest $n_\text{max}$.
As $\rho$ increases, there is a point where the quadratic and
cubic polynomial fits have equal probability.  We define
\begin{equation}
\text{AIC}(\rho, n_\text{max}) = \frac{1}{2} \chi_{5-n_\text{max}}^2 + n_\text{max} + 1
\end{equation}
and choose a third interesting value of $\rho$: \textbf{(III)}
$\text{AIC}(\rho^{(III)},2) = \text{AIC}(\rho^{(III)},3)$.  Note this does
not always have a solution, particularly if the quadratic fit has a lower AIC
at $\rho=0$. A posteriori we can rationalize this choice as the point where
the quadratic and cubic descriptions of the data are equally good (or bad)
from an information-theoretic perspective.

\begin{table}[h]
\centering
\addtolength{\tabcolsep}{3 pt}   
\begin{tabular}{l|l||l|l|l}
\multicolumn{1}{c|}{obs.} &
\multicolumn{1}{c||}{$n_\text{max}$} &
\multicolumn{1}{c|}{$\rho^{(I)}$} &
\multicolumn{1}{c|}{$\rho^{(II)}$} &
\multicolumn{1}{c}{$\rho^{(III)}$} \\
\hline\hline
$M_\pi$ & 2 & 0.0169 & 0.0158 & 0.0162 \\
        & 3 & 0.0068 & 0.0068 & 0.0162 \\
\hline
$F_\pi$ & 2 & 0.0210 & 0.0196 & 0.0197 \\
        & 3 & 0.0098 & 0.0098 & 0.0197 \\
\hline
\end{tabular}
\caption{\label{tab:rho} Various estimates of the systematic error parameter
$\rho$ as determined by methods described in the text.}
\end{table}

\subsection{Summary of Systematic Error Analysis}

From Tab.~\ref{tab:rho} we can see there are eight unique $\rho$ values
from approximately 0.007 to 0.021.  Rather than pick just one, we consider
a few summary statistics:  the arithmetic mean $\rho_\text{a} = 0.0157$,
the median $\rho_\text{m} = 0.0165$, or the geometric mean
$\rho_\text{g} = 0.0148$.  All give relatively similar values close to the central
grouping. We make a conservative choice and choose the largest of the three
$\rho = \rho_\text{m} = 0.0165$.  If we compare this estimate to the previous rough
guess of 0.01 quoted in \cite{LatticeStrongDynamics:2021gmp}, it is nice
to see they are not too different and that 0.01 falls within the range
of estimated values.  A final summary of our results with the relative systematic error
included is given in Table~\ref{tab:data_stat+sys}.

\begin{table}[ht]
  \centering
  \addtolength{\tabcolsep}{3 pt}   
  \begin{tabular}{c||l|l|l|l|l}
    $m_q$ & 0.00125 & 0.00222 & 0.00500 & 0.00750 & 0.00889 \\
    \hline
    $M_{\pi_5}$ &
    0.0811(13) & 0.1087(18) & 0.1657(27) & 0.2057(34) & 0.2253(37) \\
    $\widehat{F}_{\pi_5}$ &
    0.02168(36) & 0.02794(47) & 0.03982(66) & 0.04831(80) & 0.05262(88) \\
    $M_\sigma$ &
    0.1174(48) & 0.1545(83) & 0.2408(87) & 0.2744(85) & 0.301(12) \\
    $\widehat{F}_S$ &
    0.0254(17) & 0.0361(37) & 0.0536(33) & 0.0711(38) & 0.0787(60) \\
    $M_{a_{0,1}}$ &
    0.1536(27) & 0.2070(63) & 0.3119(58) & 0.3773(65) & 0.4193(76) \\
    $\widehat{F}_{a_{0,1}}$ &
    0.00691(17) & 0.00944(45) & 0.1480(36) & 0.01829(39) & 0.02047(54) \\
    $M_{\rho_i}$ \cite{LatticeStrongDynamics:2018hun} &
    0.1709(71) & 0.2197(52) & 0.3024(80) & 0.3696(61) & 0.4093(71)
  \end{tabular}
  \caption{\label{tab:data_stat+sys}Final summary of infinite volume ground state
  rest masses and decay constants with relative systematic error of $\rho=0.0165$
  included following Eq.~(\ref{eq:total_uncertainty}). Data for $M_{\rho_i}$
  derived from \cite{LatticeStrongDynamics:2018hun} for convenience. Results
  with only statistical errors in Table~\ref{tab:data_stat_only}.}
\end{table}

\begin{figure}[h]
    \centering
    \includegraphics[width=0.5\textwidth]{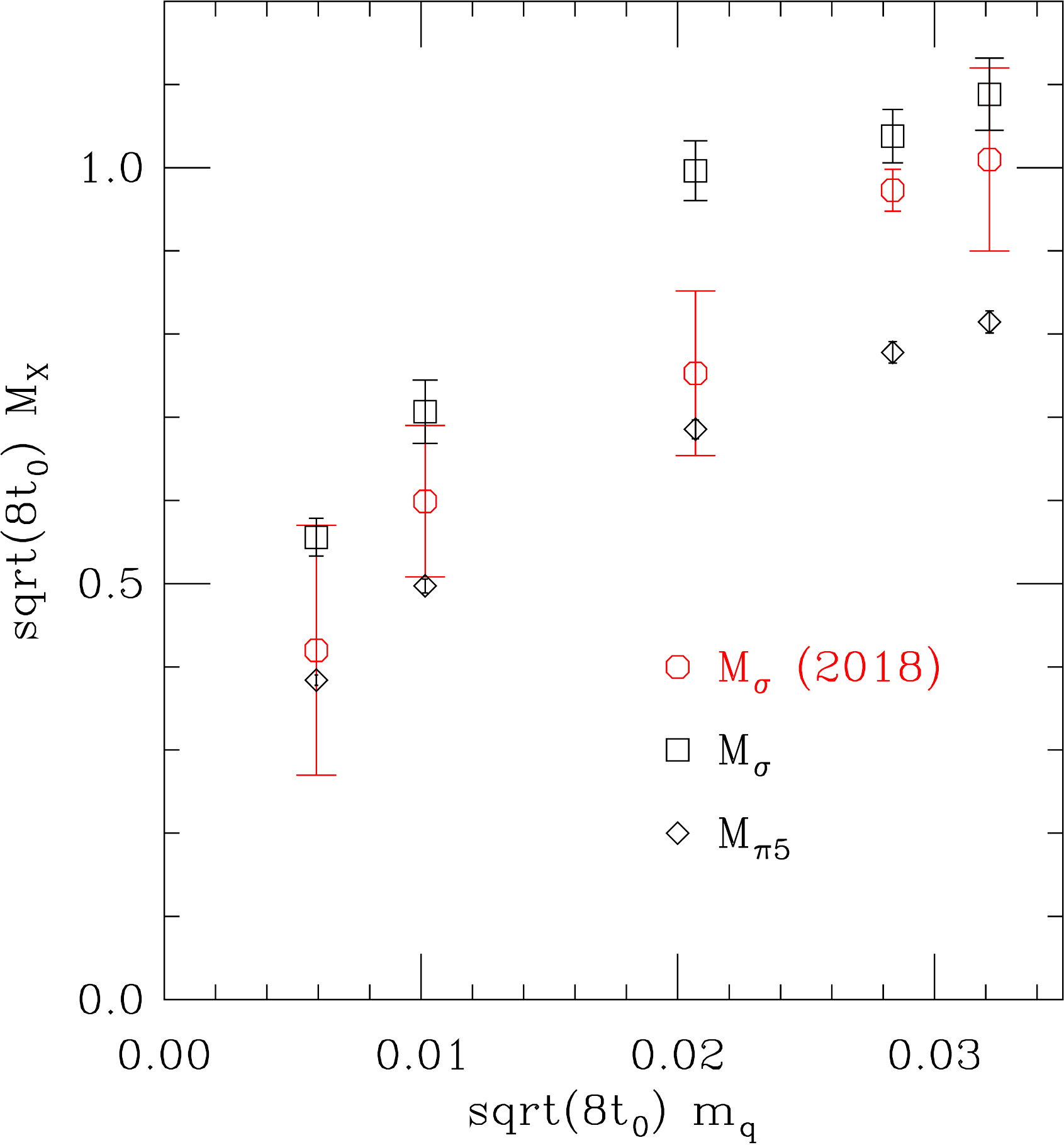}
    \caption{\label{fig:old_new} A comparison of $M_\sigma$ computed previously
    \cite{LatticeStrongDynamics:2018hun} with the results from
    Tab.~\ref{tab:data_stat+sys}.}
\end{figure}

In Fig.~\ref{fig:old_new} we compare the previously computed results
for $M_\sigma$ with combined statistical and systematic errors as described
in \cite{LatticeStrongDynamics:2018hun} with the new results
of Tab.~\ref{tab:data_stat+sys} for $M_\sigma$ and $M_{\pi_5}$.  The values
for $\sqrt{8 t_0}/a$ are taken from Tab.~I of
\cite{LatticeStrongDynamics:2018hun} and the plot style is similar to Fig.~10
of \cite{LatticeStrongDynamics:2018hun}.  The conclusion we draw from this
comparison was that the previous analysis method for computing $M_\sigma$
led to systematically lower mass values and that the method used previously
to estimate systematic errors was sufficiently conservative as to cover the
the downward shift of the result.

\section{Comments on Chiral Expansions}

While the SU(3) $N_f = 8$ theory with massive Dirac fermions is a potentially
interesting theory on its own, being a possible candidate for composite dark
matter \cite{LatticeStrongDynamics:2013elk},
the theory closer to the chiral limit $m_q \to 0$ might also be relevant for
composite Higgs phenomenology.  As stated in Sec.~\ref{sec:intro}, our results
alone are not sufficient to establish 
with certainty
whether the massless $N_f=8$ theory
is inside or outside the conformal window.  But, the low energy content
of the two scenarios is quite different: in one case a very strongly coupled
conformal field theory and in the other case massless Nambu-Goldstone bosons
and possible light flavor-singlet scalar resonance with a mass of order $F_\pi$.
Specific models will appear quite different depending on the scenario,
and when fitted to our data, those models will, in general, be an expansion
in some small parameter which vanishes in the chiral limit.  We will discuss
different specific extrapolations in detail in a companion paper
\cite{LatticeStrongDynamics:2023uzj}.

\begin{figure}[h]
\centering
\includegraphics[width=0.5\textwidth]{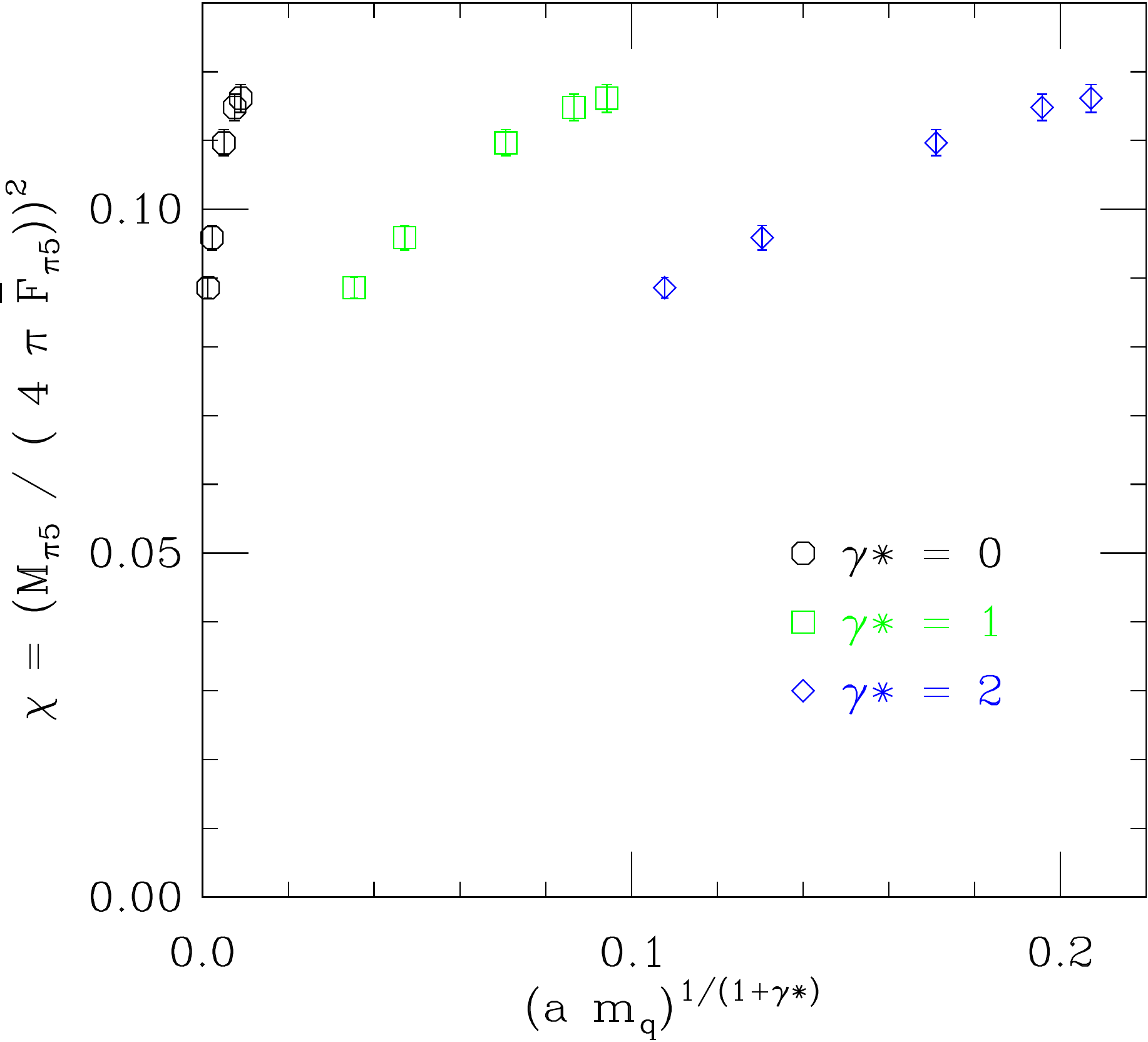}
\caption{
  Chiral expansion parameter
  $\chi = M_{\pi_5}^2 / (4 \pi \widehat{F}_{\pi_5})^2$
  \textit{vs.}\ other chiral expansion parameters $m_q^{1/(1+\gamma^*)}$.
  If the theory is conformal $\chi$ should be non-zero in the chiral limit.
  If the theory is spontaneously broken, $\chi$ should be zero in the chiral limit.
}
\label{fig:chi_vs_pow_mq}
\end{figure}

Here we note that based on one's \textit{a priori} expectation for the nature
of the low-energy theory in the chiral limit, the choice of expansion parameter
can lead to very different presentations of the data.  One could naively
plot results \textit{vs}.\ the fermion mass $m_q$, or some power of the fermion
mass $m_q^{1/(1+\gamma^*)}$, $0 \le \gamma^* \le 2$ motivated by assuming
conformal symmetry in the chiral limit,
or $\chi \equiv M_{\pi_5}^2 / (4 \pi \widehat{F}_{\pi_5})^2$ by assuming
spontaneous chiral symmetry breaking in the chiral limit.  In theory like
SU(3) $N_f = 2$ (QCD), these choices often don't make any appreciable difference
in the presentation of the data. But in this theory, if the chiral limit
is conformal, the expansion parameter $\chi$ doesn't vanish as $m_q \to 0$.
Visually, we can see the difference in Fig.~\ref{fig:chi_vs_pow_mq}.  Since the value
of $\gamma^*$ is a dynamical parameter that can only be determined through a careful
extrapolation, we plot three representative values that cover weakly and strongly-coupled
CFTs plus an intermediate value.  Regardless of which value of $\gamma^*$ is chosen,
$\chi$ varies significantly over the range of $m_q$ with a fair degree of curvature
which makes it difficult to estimate how small $m_q$ must be before the constant term
in $\chi$ dominates the leading $m_q$-dependent term.  Of course, if $\chi$ vanishes
in the chiral limit, then the constant term will never dominate. This suggests it will
be difficult to distinguish with much certainty given our current results
whether or not $\chi$ vanishes in the chiral limit.
Calculations at smaller fermion masses are needed.

\begin{figure}[h]
\centering
\includegraphics[width=0.49\textwidth]{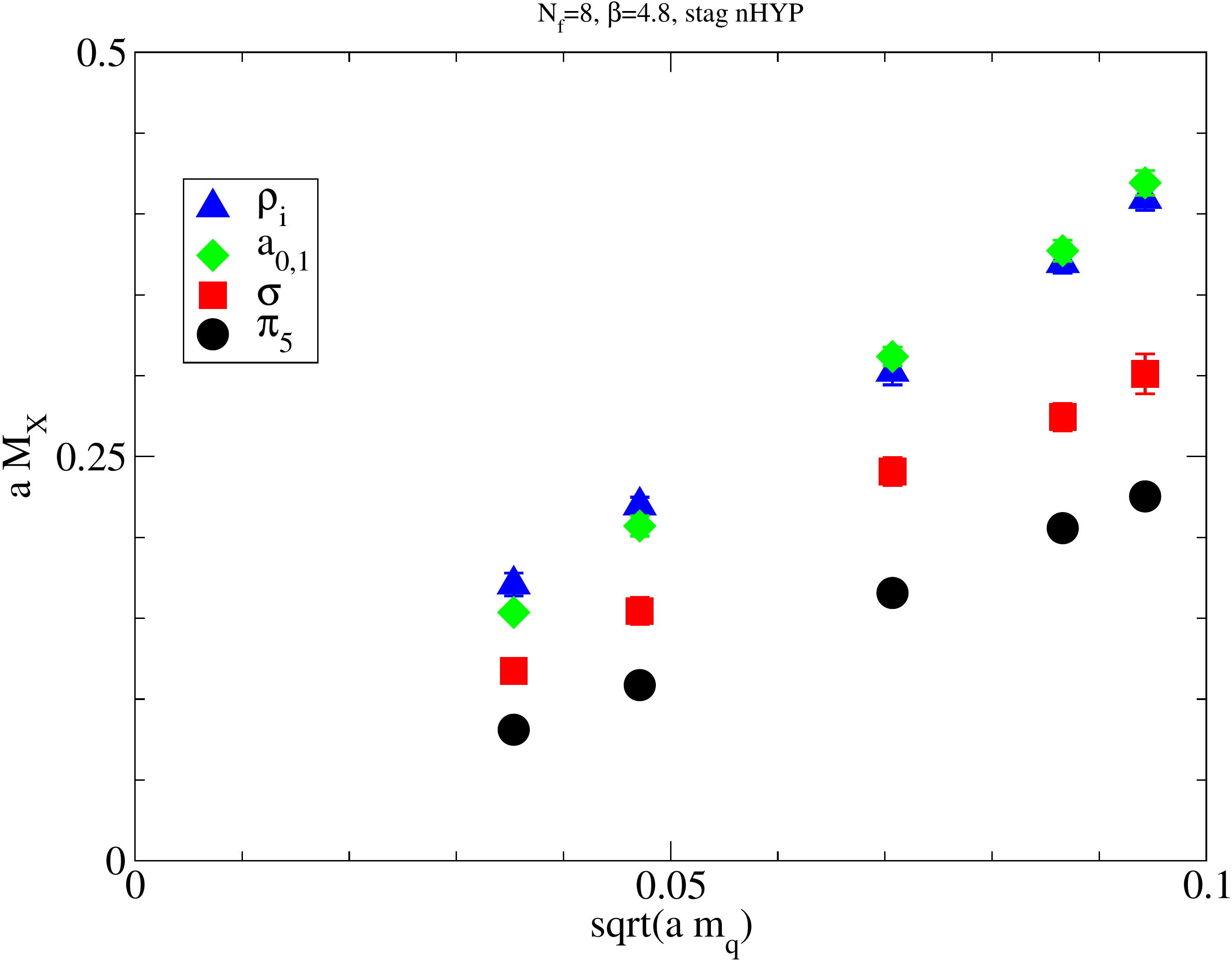}
\includegraphics[width=0.49\textwidth]{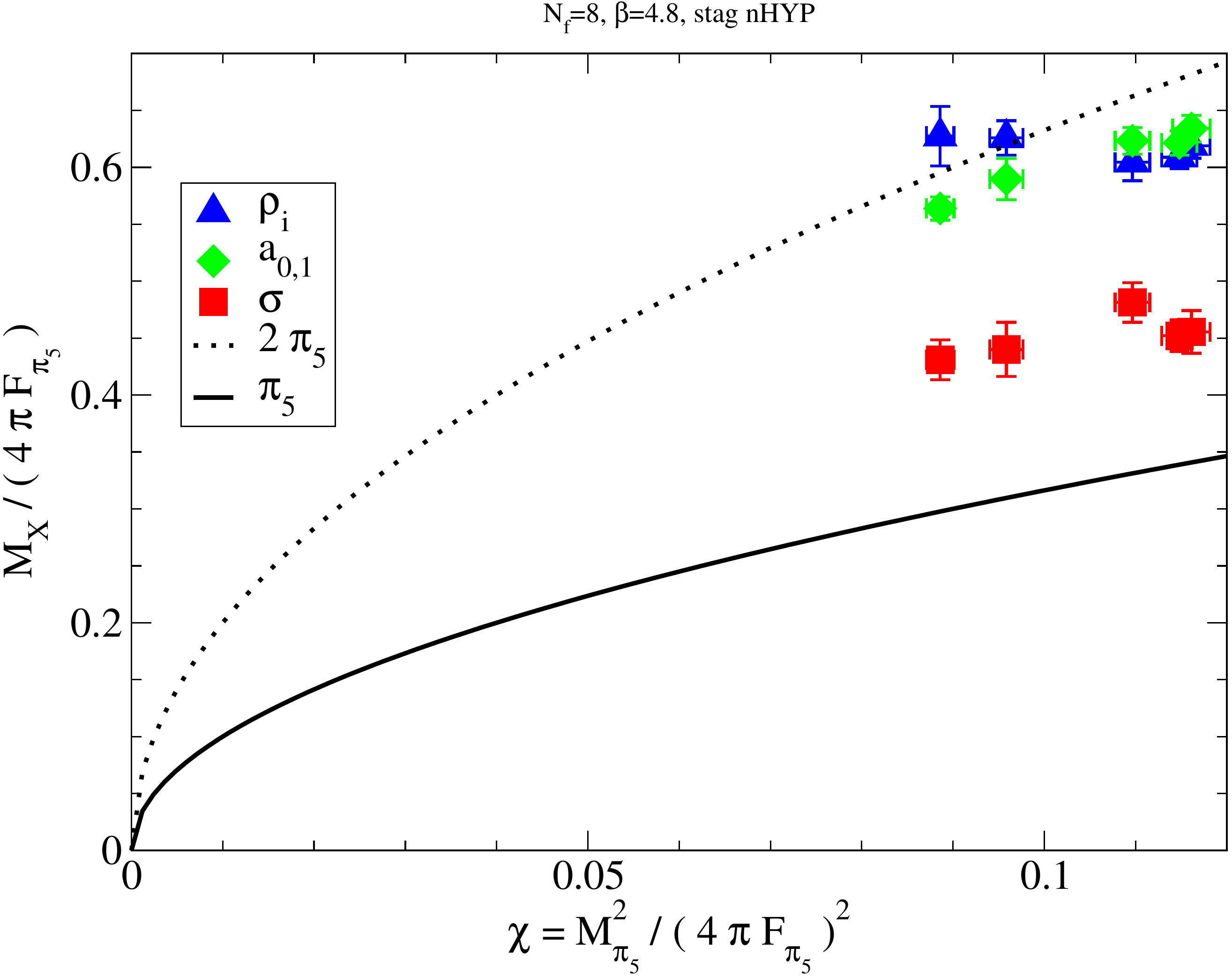}
\caption{
  Two different presentations of the spectrum from
  Tab.~\ref{tab:data_stat+sys}. On the left, in units of the lattice spacing $a$
  \textit{vs.}\ a chiral expansion parameter assuming conformal symmetry and
  $\gamma^* \approx 1$. On the right, in units of the chiral breaking scale
  $4 \pi \widehat{F}_{\pi_5}$ \textit{vs.}\ a chiral expansion parameter assuming
  spontaneous chiral symmetry breaking.  The dotted line on the right indicates
  the energy threshold for decays to two pions.
}
\label{fig:MX_vs_chiral}
\end{figure}

We can now present the results for the spectrum in two different ways.
In Fig.~\ref{fig:MX_vs_chiral}, on the left is a presentation appropriate when
assuming the theory is conformal in the chiral limit with a mass anomalous dimension
$\gamma^* \approx 1$.  In units of the lattice spacing $a$, the masses of all
the hadrons are expected to extrapolate to zero since any non-zero hadron mass
would break conformal symmetry.  On the right is a presentation assuming the chiral
symmetry is spontaneously broken in the chiral limit and the relevant scale of chiral
symmetry breaking is set by $4 \pi \widehat{F}_{\pi_5}$.  All the hadron masses except
the pion should be non-zero in the chiral limit.  Plotted this way, the pion is shown
as a simple curve since $M_{\pi_5} / 4 \pi \widehat{F}_{\pi_5} = \sqrt{\chi}$. This also
makes it easy to display the energy threshold as a dotted line for decay to two pions.
In the current data
set, both the flavor-singlet and non-singlet scalar mesons appear to be unable to
decay to two pions.

Focusing solely on the data presented in this section, it is still far from
clear whether or not the chiral parameter $\chi$ vanishes in the chiral limit.
On the other hand, $\chi$ varies significantly with the fermion mass
which also suggests we are far from the hyperscaling limit where $\chi$ should
be a non-zero constant.
Recent numerical studies with improved gauge action \cite{Hasenfratz:2022qan}
suggest that the SU(3) $N_f=8$ system could be at the opening of the conformal window,
or at least very close to it.  There are indications of an infrared fixed point
at much stronger couplings than what is probed by our data in this paper.
This is so even if $N_f=8$ is below the conformal window. Therefore, corrections to scaling
in the gauge coupling could be significant. This can explain our inability to distinguish
between the conformal and chirally broken scenarios.

\section{\label{sec:GMOR_results}Gell-Mann Oakes Renner (GMOR) Ratio Results}

As discussed in Sec.~\ref{sub:GMOR_ratio}, numerical studies of the GMOR ratio
can shed light on the low-energy behavior of the $N_f=8$ theory
by measuring how much the ground-state pion pole contributes
the pseudoscalar two-point correlation function.  A value close to unity
indicates pion pole-dominance. Tab.~\ref{tab:condensate}
shows the computed values for the chiral condensate and the integrated pseudoscalar
correlation function. Although computed by two different techniques,
the results agree extremely well with Eq.~(\ref{eq:pion_susceptibility}).
Using the largest volume data at each fermion mass for the condensate and the
statistical-only data in Tab.~\ref{tab:data_stat_only}, we compute the
GMOR ratio in Eq.~(\ref{eq:GMOR_ratio}) and propagate the statistical-only
errors.  We then apply the relative systematic error correction
estimated in Sec.~\ref{sec:systematics}. The results are shown
in the rightmost column of Tab.~\ref{tab:condensate}.  The lowest pole
does not fully dominate the pion correlation function in our fermion mass range
as the result is larger than one.  If we assume a mass-deformed CFT
is the correct low energy description, then the lowest pion pole will never
dominate the pseudoscalar correlation function at any fermion mass as it
is not a pseudo-Nambu-Goldstone boson, as in Eq.~(\ref{eq:GMOR_ratio}).

In the case the $N_f=8$ theory is outside the conformal window,
one of the poles contributing to the pion correlation function
would have a pole position at $M_{\pi_5}+M_\sigma$ and a residue
proportional to $g^2_{\pi\pi\sigma}$ which we will assume is
$\mathcal{O}(\widehat{F}_{\pi_5}^2)$.  For the leading pion pole
to dominate, $M_{\pi_5}^2 / ( M_{\pi_5} + M_\sigma )^2 \ll 1$.
In this work, the ratio varies from 0.167(8) --- 0.183(9), which
in this scenario is interpreted as not small enough to ensure
pion pole dominance.  A direct calculation of the coupling
$g_{\pi\pi\sigma}$ and/or further calculations at lighter fermion masses
should shed light on this issue.

We did not perform an infinite volume extrapolation of the condensate data
in Tab.~\ref{tab:condensate} similar to the ones described
in Sec.~\ref{sub:finite_volume}. The systematic effect of this correction might be
significant on the scale of the uncertainties shown for the GMOR ratio.
But, the effect is unlikely to be significant relative to the deviation of the ratio
from unity. In the future, if a detailed model is to be fit to these data,
the modeller should consider including these neglected corrections.

\begin{table}[h]
\centering
\addtolength{\tabcolsep}{3 pt}   
\begin{tabular}{c|c||c|c||c}
$m_q$ & $L$ & $\left\langle\overline\chi \chi \right\rangle$ &
  $\sum_t C_{\pi_5}(\vec{0},t)$ & $R_G(m_q)$ \\
\hline
0.00125 & 96 & 0.0121704(53) & 9.732(28) & 2.462(42) \\
        & 64 & 0.0121220(70) & 9.641(17) & \\
\hline
0.00222 & 48 & 0.019918(14)  & 9.039(25) & 2.397(45) \\
\hline
0.00500 & 48 & 0.040808(18)  & 8.164(41) & 2.344(41) \\
        & 32 & 0.040521(21)  & 8.103(12) & \\
\hline
0.00750 & 48 & 0.058447(12)  & 7.808(20) & 2.219(37) \\
        & 32 & 0.058445(26)  & 7.786(16) & \\
        & 24 & 0.058063(73)  & 7.752(32) & \\
\hline
0.00889 & 32 & 0.068086(28)  & 7.661(37) & 2.152(38) \\
        & 24 & 0.067814(46)  & 7.615(17) & \\
\end{tabular}
\caption{Values for the staggered chiral condensate $\left\langle\overline\chi \chi
\right\rangle$, computed using a noisy estimator, and the integrated pseudoscalar
correlation function, computed using a point source.  The reader can verify
that the columns satisfy Eq.~(\ref{eq:pion_susceptibility}). Only statistical
errors are shown. The rightmost column shows the GMOR ratio defined
in Eq.~(\ref{eq:GMOR_ratio}) with errors computed as described in the text.}
\label{tab:condensate}
\end{table}

\section{Discussion}

In this investigation, we have made many methodological improvements
with respect to our earlier lattice study of the
$N_f=8$ theory \cite{LatticeStrongDynamics:2018hun}.
In particular, we have employed two
different methods for dealing with time-independent contributions
to the flavor-singlet scalar correlator, first by using the subtraction scheme developed
in Sec.~\ref{sub:2pt_const} and then by working with moving frames and applying
the dispersion relation described in Sec.~\ref{sub:dispersion_relation}.  We were able
to substantially reduce the systematic uncertainties of our fit results
using the Bayesian model
averaging approach.  Additionally, we used improved ``linear'' shrinkage estimators
for data covariance which we found were more reliable given the amount of statistics.
There was an open question in our previous paper whether finite volume effects
could be significant even when $M_{\pi_5} L \gtrsim 5.3$.  Now, we can
see that the finite volume effects are mild and don't play
a significant role in the final result.  We find that $M_\sigma / M_{\pi_5}$ ranges
from 1.45 -- 1.34 as $M_{\pi_5} / M_{\rho_i}$ increases from 0.47 -- 0.55.

We computed a new observable, the scalar decay
constant $\widehat{F}_S$, which, as we show in a related paper
\cite{LatticeStrongDynamics:2023uzj}, provides useful
independent constraints on various low-energy effective theories. We also computed
the flavor-nonsiglet scalar meson mass $M_{a_{0,1}}$ and decay constant
$\widehat{F}_{a_{0,1}}$.  The proximity of the $a_{0,1}$ to the decay threshold
suggests that a careful elastic scattering analysis might be warranted in the future
if more accurate results are desired.

\begin{acknowledgments}
  R.C.B.~and C.R.~acknowledge United States Department of Energy (DOE) Award No.~{DE-SC0015845}.
  K.C.~acknowledges support from the DOE through the Computational Sciences Graduate Fellowship
  (DOE CSGF) through grant No.~{DE-SC0019323} and also from the P.E.O. Scholar award.
  G.T.F.~acknowledges support from DOE Award No.~{DE-SC0019061}.
  A.D.G.~is supported by SNSF grant No.~{200021\_17576}.
  A.H.~and E.T.N.~acknowledge support by DOE Award No.~{DE-SC0010005}.
  J.I. acknowledges support from ERC grant No. 101039756.
  D.S.~was supported by UK Research and Innovation Future Leader Fellowship {MR/S015418/1} and STFC grant {ST/T000988/1}.
  P.V.~acknowledges the support of the DOE under contract No.~{DE-AC52-07NA27344} (Lawrence Livermore National Laboratory, LLNL).

  We thank the LLNL Multiprogrammatic and Institutional Computing program for Grand Challenge supercomputing allocations. We also thank Argonne Leadership Computing Facility (ALCF) for allocations through the INCITE program. ALCF is supported by DOE contract No.~{DE-AC02-06CH11357}. Computations for this work were carried out in part on facilities of the USQCD Collaboration, which are funded by the Office of Science of the DOE, and on Boston University computers at the MGHPCC, in part funded by the National Science Foundation (award No.~{OCI-1229059}). This research utilized the NVIDIA GPU accelerated Summit supercomputer at Oak Ridge Leadership Computing Facility at the Oak Ridge National Laboratory, which is supported by the DOE Office of Science under Contract No.~{DE-AC05-00OR22725}.
\end{acknowledgments}

\appendix

\section{\label{app:ensembles}Ensembles}

A summary of the ensembles used in this paper are shown in Tables
\ref{tab:ensembles_heavy} and \ref{tab:ensembles_light}.

\begin{table}[h]
\centering
\begin{tabular}{c|c|c|c|c|c|c}
Volume & Mass & Try & MDTU & Period (MDTU) & Block (MDTU) & $N_\text{blk}$ \\
\hline\hline
$24^3 \times 48$ & 0.00889 & 1 & [250,25000] & 10 & \textbf{100} & \textbf{247} \\
\hline\hline
$32^3 \times 64$ & & 1 & [1040,7000] & 40 & 80 & 75 \\
& & 2 & [1040,7000] & 40 & 80 & 75 \\
& & 3 & [1040,7000] & 40 & 80 & 75 \\
& & 4 & [1040,7000] & 40 & 80 & 75 \\
\hline
& & \textbf{C} & & & \textbf{80} & \textbf{300} \\
\hline\hline
$24^3 \times 48$ & 0.0075 & 1 & [350,10000] & 10 & \textbf{90} & \textbf{107} \\
\hline\hline
$32^3 \times 64$ & & 1 & [255,1395] & 10 & 100 & \\
& & & [1400,25160] & 5 & 100 & \textbf{249} \\
\hline\hline
$48^3 \times 96$ & & 1 & [250,9990] & 10 & 70 & 139 \\
& & 2 & [250,9990] & 10 & 70 & 139 \\
\hline
& & \textbf{C} & & & \textbf{70} & \textbf{278} \\
\hline\hline
$32^3 \times 64$ & 0.005 & 1 & [251,29641] & 5 & 100 & 293 \\
& & 2 & [20011,22815] & 2 & 100 & 28 \\
& & 3 & [29001,31653] & 2 & 100 & 26 \\
& & 4 & [10001,13293] & 2 & 100 & 32 \\
\hline
& & \textbf{C} & & & \textbf{100} & \textbf{379} \\
\hline\hline
$48^3 \times 96$ & & 1 & [250,4200] & 10 & 50 & 79 \\
& & 2 & [250,3390] & 10 & 50 & 63 \\
\hline
& & \textbf{C} & & & \textbf{50} & \textbf{142} \\
\hline\hline
\end{tabular}
\caption{\label{tab:ensembles_heavy}
Ensembles, or Markov chains, used in this study with
$0.005 \le m_q \le 0.00889$.
``Try'' assigns a label to each Markov chain and the label ``\textbf{C}'' indicates
the combined summary for all chains at a given mass and volume. ``Period'' indicates
how often the correlation functions were computed.
}
\end{table}

\begin{table}[h]
\centering
\begin{tabular}{c|c|c|c|c|c|c}
Volume & Mass & Try & MDTU & Period (MDTU) & Block (MDTU) & $N_\text{blk}$ \\
\hline\hline
$48^3 \times 96$ & 0.00222 & 1 & [250,11190] & 2 & 120 & 91 \\
& & 2 & [1000,9930] & 2 & 120 & 74 \\
& & 3 & [210,1450] & 10 & 120 & 10 \\
& & 4 & [210,1410] & 10 & 120 & 10 \\
& & 5 & [210,1360] & 10 & 120 & 9 \\
& & 6 & [210,1290] & 10 & 120 & 9 \\
& & 7 & [210,1350] & 10 & 120 & 9 \\
\hline
& & \textbf{C} & & & \textbf{120} & \textbf{212} \\
\hline\hline
$64^3 \times 128$ & 0.00125 & r0 & [200,2060] & 10 & 120 & 15 \\
& & r1 & [200,1990] & 10 & 120 & 15 \\
& & r2 & [200,2010] & 10 & 120 & 15 \\
& & r3 & [200,2070] & 10 & 120 & 15 \\
& & s0 & [8436,17088] & 12 & 120 & 72 \\
& & s1 & [7644,17472] & 12 & 120 & 82 \\
& & s2 & [7212,17412] & 12 & 120 & 86 \\
\hline
& & \textbf{C} & & & \textbf{120} & \textbf{300} \\
\hline\hline
$96^3 \times 128$ & & 2 & [500,3144] & 2 & 80 & 34 \\
& & 3 & [500,3282] & 2 & 80 & 35 \\
\hline
& & \textbf{C} & & & \textbf{80} & \textbf{69} \\
\hline\hline
\end{tabular}
\caption{\label{tab:ensembles_light}
Ensembles, or Markov chains, used in this study with
$0.00125 \le m_q \le 0.00222$.
``Try'' assigns a label to each Markov chain and the label ``\textbf{C}'' indicates
the combined summary for all chains at a given mass and volume. ``Period'' indicates
how often the correlation functions were computed.
}
\end{table}

\section{\label{sec:prob_norm}Normalization of model probabilities}

When performing an aggressive model averaging analysis by considering a wide range of models $\{M\}$
and a wide range of data subset selections $T_1$ for each model, the resulting set of $\log p(M|D)$
can vary by several orders of magnitude, making it numerically challenging to perform an accurate calculation
of $\sum_{\{M\}} p(M|D)$.  In particular, exponentiating each $\log p(M|D)$ and then performing the sum seems
like a bad idea. So, we work directly with $\log p(M|D)$ to compute the log of the sum.
Let $\ell_n$ be a sorted list of the $\log p(M|D)$: $\ell_1 \le \ell_2 \le \cdots \le \ell_N$.
We can construct the partial sums recursively
\begin{equation}
s_1 = \ell_1 , \qquad s_{n} = s_{n-1} + \log \left( 1 + e^{\ell_n - s_{n-1}} \right) \quad (n>1)
\end{equation}
The final sum over model probabilities is  $\sum_{\{M\}} p(M|D) = \exp s_N$.
The key observation is that sorting the list ensures that two wildly different numbers
are not combined at any step with accompanying large loss of precision.

\section{\label{sec:VarCov} Unbiased Sample Estimator for the Variance of the Covariance}

Using Mathematica's \texttt{MomentConvert[]} functionality, it is a few lines of code
to express the unbiased sample estimator for $\widehat{\Var}(\Sigma_{ij})$ in terms of raw moments
\begin{verbatim}
centMom11Est = MomentConvert[CentralMoment[{1, 1}], "SampleEstimator"];
bias = MomentConvert[centMom11Est, {Moment, n}];
MomentConvert[(centMom11Est - bias)^2, {Moment, n}]
\end{verbatim}
The result is
\begin{eqnarray}
\lefteqn{\frac{n^3}{n-1} \widehat{\Var}(\Cov(x,y))} \nonumber \\*
& = & (n-1)\mu_{2,2} - 2 (n-1) \left( \mu_{2,1} \mu_{0,1} + \mu_{1,0} \mu_{1,2} \right) 
+ \mu_{2,0} \mu_{0,2}
+ (n-2)\left( \mu_{2,0} \mu_{0,1}^2 + \mu_{1,0}^2 \mu_{0,2} \right) \nonumber \\*
&& - (n-2) \mu_{1,1}^2 +2(3n-4)\mu_{1,1}\mu_{1,0}\mu_{0,1} 
- 2(2n-3)\mu_{1,0}^2 \mu_{0,1}^2
\end{eqnarray}
where we use Mathematica's convention for raw moments
\begin{equation}
\mu_{i,j;\mathcal{S}} = \frac{1}{n} \sum_{(x,y)\in\mathcal{S}} x^i y^j 
\end{equation}
Following P{\'e}bay \cite{Pebay:2008}, we would like to construct a one-pass, parallelizable computation.
To explain the notation, $\mathcal{S}$ is a set of $n$ samples that can be partitioned
into two subsets $\mathcal{S}_1$ and $\mathcal{S}_2$ of $n_1$ and $n_2$ samples, respectively,
so $n_1 + n_2 = n$.  The computation can be parallelized
by performing computations on the subsets and combining the results.
In the special case where $n_1 = n - 1$ and $n_2 = 1$, the results simplify and can be used as a
one-pass algorithm
\begin{equation}
\mu_{i,j;\mathcal{S}} = \frac{n_1}{n} \mu_{i,j;\mathcal{S}_1} + \frac{n_2}{n} \mu_{i,j;\mathcal{S}_2} 
= \mu_{i,j;\mathcal{S}_1}
+ \frac{n_2}{n} \left( \mu_{i,j;\mathcal{S}_2} - \mu_{i,j;\mathcal{S}_1} \right)
\end{equation}
where the first form is symmetric and more useful when $\mathcal{S}_1$ and $\mathcal{S}_2$
are of comparable size and the second form is better suited when $\mathcal{S}_2$ is a single
sample $(x,y)$
\begin{equation}
\mu_{i,j;\mathcal{S}}
= \mu_{i,j;\mathcal{S}_1} + \frac{1}{n} \left( x^i y^j - \mu_{i,j;\mathcal{S}_1} \right) .
\end{equation}

\bibliography{main}

\end{document}